\documentstyle[epsf,aps]{revtex}  

\begin{document}
\draft
\title{
Hybrid stars : Spin polarised nuclear matter and density dependent
quark masses 
}
\vskip 1.0 true cm
\author
{V.S. Uma Maheswari$^{1,3}$, J.N. De$^1$
and S.K. Samaddar$^2$ 
}
\vskip 0.7 true cm
\address
{
$^1$ Variable Energy Cyclotron Centre, \\
1/AF, Bidhan Nagar, Calcutta - 700064, India\\
$^2$ Saha Institute of Nuclear Physics, \\
1/AF, Bidhan Nagar, Calcutta - 700064, India\\
$^3$Institut f\"ur Theoretische Physik der Universit\"at T\"ubingen,\\
Auf der Morgenstelle 14, D-72076 T\"ubingen, Germany
}
\maketitle  
\vskip 1.0 true cm
\begin{abstract}
The possibility of formation of a droplet phase (DP) inside a
star and its consequences on the structural properties
of the star are investigated. For nuclear matter (NM), an
equation of state (EOS) based on finite range, momentum 
and density dependent (FRMDD) interaction, and which 
predicts that neutron matter undergoes ferromagnetic 
transition at densities realisable inside the neutron 
star is employed. An EOS for quark matter (QM) with 
density dependent quark masses, the so-called 
effective mass model, is constructed by correctly
treating the quark chemical potentials.
A comparative study of hybrid star properties as obtained
within the usual bag model and the effective mass model
shows that both these models yield similar results.
Then, the effect of spin polarisation on the formation 
of DP is investigated. 
Using the EOS based
on FRMDD interaction alongwith the usual bag model, 
it is also found that a droplet phase consisting of 
strange quark matter and unpolarised nuclear matter
sandwiched between a core of
polarised nuclear matter and
a crust containing unpolarised nuclear matter
exists. 
Moreover, one could, in principle, 
explain the mass and surface magnetic field
satisfactorily, and as well allow, 
due to the presence of a droplet phase,
the direct URCA process to happen.
\vspace{5mm}\\
{\em Keywords: }Hybrid stars, Ferromagnetic phase in nuclear
matter, Effective quark masses, Droplet phase
\end{abstract}
\pacs{97.60.Jd; 21.65.+f}
\section{Introduction}

Since the discovery of pulsars \cite{hewish} and their identification
 \cite{gold} as rotating neutron stars during the late 1960's, much
theoretical work have been done to study the structure of neutron stars.
At the core of these objects, the density could be as high as five to ten
times the nuclear matter density ($\rho_o \simeq 0.16\ fm^{-3}$) which falls
down by several orders of magnitude in the crust region.
The nature of matter at such high densities in the core is still an
unresolved issue \cite{prakash}. Over the two decades, many of the
studies were addressed towards a resolution of this problem
(see for e.g. \cite{prakash,shapiro}).

It has been conjectured in particular that there exists a
phase transition between nuclear matter (NM) and quark matter (QM)
at high densities [5-11]. Witten \cite{witten} has further
speculated that the strange quark matter (SQM) consisting of $u$,
$d$ and $s$ quarks may be the absolute ground state of hadronic
matter.
If this is true, then the possibility that pulsars are objects
made up of purely SQM cannot be ruled out.
Recently, from a semiempirical analysis of the mass-radius
relationship Li {\it et al } \cite{li} have suggested that Her X-1
pulsar maybe a strange quark star (see also \cite{madsen}).
On the basis of these hypotheses, several authors [3,5-11] have
studied the properties of stars
consisting of only SQM, {\it i.e.} strange quark stars (SQS)
and neutron stars having quark cores with
nucleon envelopes, {\it viz.} hybrid stars (HS).
In general, the transition between the nuclear and quark 
matter is considered to occur at a unique pressure determined
by Gibbs criteria, ( we consider the transition to be first order).
Inside the star, the pressure decreases smoothly outwards. The 
mixed phase has constant pressure independent of the volume 
fraction of the phases and therefore, the mixed phase collapses
to a single point on the density profile of the star.
Consequently, the mixed phase cannot exist in the hybrid star.
Glendenning \cite{glen} has recently proposed that for systems
with more than one conserved charge, the QM and NM could coexist
for a finite range of pressures. A well-defined mixed phase may
therefore be present \cite{glen,heisel,drago}
 over a finite region within the star
(hereafter referred to as the droplet phase ).

Occurrence of a droplet phase (DP) have quite important observational
consequences. In particular, the values of $\beta-$equilibriated 
proton fraction $x_p$ in DP can be large enough to allow direct URCA 
process to happen, and hence, is of significance to the cooling process 
of neutron stars. However, whether or not such a DP is energetically
favourable depends largely upon the nature of dense matter.
As yet, the exact nature of dense nuclear matter 
is not known, and several 
possibilities are equally probable.
Very recently, we had shown \cite{Um97}, 
using a well-defined, finite range, 
momentum and density dependent interaction, that pure neutron matter
undergoes a ferromagnetic phase transition at density 
$\rho \sim 4\rho_{0}$.
Using such an equation of state, we then made a detailed study of
structural properties of neutron and hybrid stars, and showed
that one could, in principle, explain both the surface magnetic field
and the standard structural properties.
An interesting result is that $\beta-$equilibriated
proton fraction $x_p$ decreases sharply once the
star matter becomes spin polarised (see Fig. 5 of Ref. \cite{Um97}).
One then naturally wonders, what is the effect of such phase 
transitions on the occurrence and properties of DP ?
In particular, whether or not ferromagnetic domains, {\it i.e.} 
droplets of spin polarised matter, exist inside a star. 
And, what is the role of spin polarisation in the determination of
$x_p$, and thereby, on the possibility of direct URCA process.
In view of these, we mainly investigate the following.
Firstly, whether the proposed equation of state, based on an interaction
with firm basis in the well known properties of nuclear matter
and finite nuclei, which predicts a ferromagnetic phase transition 
permits a droplet phase. Secondly, if such a phase is allowed, what are
the effects of spin polarisation on the star structure ?
Finally, can one consistently describe the standard structural properties 
and surface magnetic field, as well as find the finite probability
 for the direct 
URCA process to happen ?

Now, in addition to the properties of dense nuclear matter, 
the nature of the quark matter equation of state can also affect
the possibility of formation of a droplet phase.
The theoretical framework that has been used in general to
study HS and SQS is the MIT bag model \cite{chodos}.
In this model, the non-perturbative aspects of quark-quark
interaction, {\it i.e.} the confinement property is invoked
through a bag parameter.
An alternate way of implementing confinement has also been
explored \cite{olive,fowler,boal}, 
which is hereafter referred to as effective
mass model of confinement. 
In this approach, the quark masses are assumed
to be density dependent. In the limit of high densities, their
values tend towards the current quark masses; in the limit of
zero density, the quark masses become infinite.
Attempts have been made \cite{chakra,ben} to study the properties
of SQS in this approach. In Ref. \cite{chakra}, it was noted that
the properties of SQM obtained using the density dependent quark
masses are quite different from those obtained using the MIT bag
model. This was subsequently contested in Ref. \cite{ben}, where
it was found that both the approaches yield similar results.
An intriguing aspect in the latter study is that the pressure
does not vanish at the density where the energy per baryon has a
minimum. This contradicts the very definition of the ground
state of a system. 
As shown in the following section, both these studies have 
incorrectly determined the quark chemical potentials. Due 
to the explicit density dependence of quark masses, there is 
an extra term, the so-called rearrangement energy, in the 
definition of chemical potential.
Taking this term into account, it is shown
 that the energy minimum and zero pressure occur at the
same density point. 
Further,  we make a comparative study
of properties of hybrid stars as obtained 
using the effective mass model 
and bag model of quark confinement. 

It may also be said here that in addition to the bulk properties
of the equations of state, the finite size effects 
 \cite{heisel,glenf} such as Coulomb, 
surface and curvature contributions can affect the possible formation of 
the droplet phase inside a star.
However, it is known that surface tension of QM 
is poorly determined \cite{heisel}, 
and is also found to be dependent upon the shape of the confining
potential \cite{Um96}.
Moreover, the internal structure of the
droplet phase is generally determined by the
competition between the Coulomb and surface energies, and
can adopt shapes like rods, plates and spheres \cite{heisel}.
Here, since we are mainly concerned with the effect of spin
polarisation in the formation of the droplet phase, 
 the finite size effects are presently not
considered.

\section{Theoretical framework for quark matter EOS }

In the following, we briefly outline the procedure to obtain the
quark equation of state in the effective mass approximation.

\subsection{The effective mass approximation }

The two important properties of QCD, namely the asymptotic freedom and
the confinement, are adequately represented by considering that quarks
are bound by a local Dirac-scalar potential, $S({\vec r}) \sim
c_l { \mid {\vec r} \mid }^l $ \cite{hooft,tegen}.
Presence of a strong Dirac-scalar component is found to be consistent
with the concept of chiral symmetry \cite{nambu}.
In terms of $ S( {\vec r})$,
one can then define the quark effective mass as
$m_q( {\vec r}) = m_q^0 + c_l { \mid {\vec r} 
\mid }^l $ \cite{tegen,parija}.
Its average value in the bulk matter limit is taken to be of the form,
\begin{equation}
m_q = <m_q({\vec r})> \equiv m_q^0+C\rho_b^{-n} \quad ; \quad n=l/3 >0,
\label{effm}
\end{equation}
where $m_q^0$ is the current quark mass, $\rho_b$ is the baryon number
density and $C$ and $n$ are parameters to be determined by requiring that
strange quark matter is stable or unstable against the normal nuclear matter.
In this effective mass model, quark-quark interactions are
assumed to be included through the density dependence of the
quark masses. Further, Eq. (\ref{effm}) implies that in the asymptotically
free regime, {\it i.e.} at high densities, $m_q \longrightarrow m_q^0$,
the current quark mass. In the limit $\rho_b \longrightarrow 0$,
$m_q \longrightarrow \infty $; this gives rise to absolute confinement.

\subsection{Rearrangement energy and Hugenholtz-Van Hove theorem }

Consider a system of one kind of quarks with an effective mass $m_q$
defined in Eq. (\ref{effm}). 
Then, the total quark number $N_q$ and the energy
$E_q$ of the given system are determined using the expressions:
\begin{eqnarray}
N_q &=& \sum_k n(k), \nonumber \\
E_q &=& \sum_k n(k) {\sqrt {k^2+m_q^2} },
\label{qnum}
\end{eqnarray}
where $k$ is the momentum and
$m_q$ is in units of $\hbar c$, with $\hbar=c=1$. The occupation
probability function $n(k)$, at a given temperature $T$ is
\begin{equation}
n(k)=
{\left [ 1+exp \left ( {\epsilon_k -\mu_q \over T} \right ) \right ]}^{-1},
\end{equation}
where $\mu_q$ is the quark chemical potential. The single particle energies
$\epsilon_k$ are determined using the standard definition,
$\epsilon_k = \partial E_q/\partial n(k) $.
In the limit of zero temperature, one gets,
\begin{eqnarray}
\epsilon_k & = & { \sqrt {k^2 +m_q^2}}+{g\over 4\pi^2}\
{\partial m_q\over \partial \rho_q } \large [ m_qk_F \sqrt {k_F^2+m_q^2}-
\nonumber \\
\quad & \quad & m_q^3\ ln \left ( {k_F +\sqrt{k_F^2+m_q^2}\over m_q } \right )
\large ], \nonumber \\
\quad & \equiv & \sqrt { k^2+m_q^2} + U(\rho_q ),
\label{spen}
\end{eqnarray}
where $\rho_q(=3\rho_b)$ is the quark number density and the
spin-colour degeneracy factor $g=6$. The term $U(\rho_q )$ in Eq. (\ref{spen})
arises solely due to the density dependence of quark mass. In the context
of nuclear physics, such a term is normally referred to as the
rearrangement energy \cite{bohr,db}.
Because of this term, the single particle levels
depend explicitly upon the value of the Fermi momentum $k_F$.
In the limit $T\longrightarrow 0$, one can obtain from Eq. (\ref{spen}) the
chemical potential $\mu_q$, equivalently the Fermi energy $\epsilon_F$
 to be
\begin{equation}
\epsilon_F \equiv \mu_q \mid_{T=0} = \sqrt { k_F^2+m_q^2} + U(\rho_q ).
\label{epsif}
\end{equation}
Similarly, the quark number density $\rho_q = N_q/V$ and the energy
density $\epsilon_q=E_q/V$ are determined using Eq. (\ref{qnum}) as
\begin{eqnarray}
\rho_q &=& {gk_F^3\over 6\pi^2 }, \nonumber \\
\epsilon_q &=& {g\over 16\pi^2}
\large [ (2k_F^3+m_q^2k_F) \sqrt {k_F^2+m_q^2}-
\nonumber \\ \quad & \quad &
m_q^4\ ln \left ( {k_F +\sqrt{k_F^2+m_q^2}\over m_q } \right )
\large ],
\label{rhoq}
\end{eqnarray}
where $V$ is the total volume of the system. In the limit
$C\longrightarrow 0$, $U(\rho_q ) \longrightarrow 0$, and one
arrives at the usual Fermi gas model expressions for $\mu_q$
and $\epsilon_q$. Starting with the standard definition (\ref{qnum})
for energy, one thus arrives at a new feature in the expression
for $\mu_q$ in the effective mass approximation as compared to
the bag model.

We now demonstrate that Eqs. (\ref{epsif}) and (\ref{rhoq}) 
 are consistent with the
well-known Hugenholtz-Van Hove (HVH) theorem \cite{hugen} arrived
at in the context of interacting Fermi systems.
The HVH theorem in general deals with the single particle
properties of an interacting Fermi gas at $T=0$. For a system
of one kind of particles, the theorem states that
\begin{equation}
\mu_q = \epsilon_F=
{ \left ( {\partial E_q\over \partial N_q } \right ) }_V.
\label{hvh}
\end{equation}
It is straightforward to see that
${ ( {\partial E_q / \partial N_q } ) }_V = E_q/N_q+\rho_q
{ [ {\partial (E_q/N_q) / \partial \rho_q } ] }_V$.
Then at equilibrium, {\it i.e.} at the density corresponding
to zero pressure one has as a special case, $\mu_q = E_q/N_q$.
Thus, it relates the quark chemical potential at equilibrium
to the average energy per particle.

Using Eqs. (\ref{rhoq}) and (\ref{hvh}), we then obtain,
\begin{eqnarray}
\mu_q &=& { \left ( {\partial E_q\over \partial N_q } \right ) }_V
,\nonumber \\
\quad &=& {\partial \epsilon_q \over \partial \rho_q }
,\nonumber \\
\quad &=& \sqrt { k_F^2 + m_q^2 } + U(\rho_q )
,\end{eqnarray}
which is same as in Eq. (\ref{epsif}) demonstrating the consistency with the
HVH theorem. The results are equally true for multicomponent
quark systems also.
We do this exercise in order to emphasize
that earlier studies \cite{chakra,ben} do not
include such a rearrangement term in their definitions of $\mu_q$.

\subsection{The equation of state }

Here, we construct, within the effective mass approximation,
the equation of state of
$\beta-$ equilibrated, electrically neutral
quark matter.

The total kinetic energy density of a
system of non-interacting, relativistic
quarks of flavour $\tau $ and
effective mass $m_{\tau }$ is given as,
\begin{equation}
\epsilon_{\tau } = {3\over 8\pi^2 }\ {{(m_{\tau }c^2)}^4\over {(\hbar c)}^3}
\left [ x_{\tau }{\sqrt {1+x_{\tau}^2}}\ (1+2x_{\tau }^2) -
ln(x_{\tau} +{\sqrt {1+x_{\tau}^2}}) \right ],
\label{etau}
\end{equation}
where $x_{\tau }=p_F^{\tau}/(m_{\tau}c)$, $p_F^{\tau}$ is
the  Fermi
momentum and is related to the quark number density $\rho_{\tau }$ of
a given flavour as $p_F^{\tau} = \hbar {(\pi^2 \rho_{\tau})}^{1/3}$.
The densities pertaining to the three flavours
can be expressed in terms of the
total quark number density $\rho_q$ and the asymmetry parameters
$\delta_{ud}$ and $\delta_{us}$ as:
\begin{eqnarray}
\rho_u &=& (\rho_q/3) \left [ 1-\delta_{ud}-\delta_{us} \right ]
\equiv (\rho_q /3 ) f_u, \nonumber \\
\rho_d &=& (\rho_q/3) \left [ 1+2\delta_{ud}-\delta_{us} \right ]
\equiv (\rho_q /3 ) f_d, \nonumber \\
\rho_s &=& (\rho_q/3) \left [ 1-\delta_{ud}+2\delta_{us} \right ]
\equiv (\rho_q /3 ) f_s,
\label{rhoi}
\end{eqnarray}
where $\delta_{ud} = (\rho_d-\rho_u)/\rho_q$,
$\delta_{us} = (\rho_s-\rho_u)/\rho_q$ and
$\rho_q=\rho_u+\rho_d+\rho_s=3\rho_b$.
Similarly, the energy density $\epsilon_L$
pertaining to a system of
relativistic non-interacting electron gas is obtained as
\begin{equation}
\epsilon_{L } = {1\over 8\pi^2 }\ {{(m_{e}c^2)}^4\over {(\hbar c)}^3}
\left [ x_{e}{\sqrt {1+x_{e}^2}}\ (1+2x_{e}^2) -
ln(x_{e} +{\sqrt {1+x_{e}^2}}) \right ],
\end{equation}
where $x_{e}=p_{F,e}/(m_{e}c)$, $m_e=0.511$ MeV,
and $p_{F,e}$ is the electron Fermi
momentum related to the density $\rho_{e}$
 as $p_{F,e} = \hbar {(3 \pi^2 \rho_{e})}^{1/3}$.

The equilibrium composition of
the quark matter is then determined subject to the $\beta-$
equilibrium conditions,
\begin{equation}
\mu_d - \mu_u = \mu_e \quad {\rm and} \quad \mu_d = \mu_s,
\label{qbeta}
\end{equation}
and the charge neutrality condition,
\begin{equation}
\rho_e = {1\over 3} \left ( 2\rho_u - \rho_d - \rho_s \right ).
\label{qcharge}
\end{equation}
Using Eq. (\ref{rhoi}) one obtains,
$\rho_e = -(\rho_q/3)\ (\delta_{ud}+\delta_{us})$.
Similarly, one can express the chemical potentials $\mu_u$, $\mu_d$
and $\mu_s$ in terms of the three quantities $\rho_q$,
$\delta_{ud}$ and $\delta_{us}$ as follows:
\begin{eqnarray}
\mu_u &=& {\left ( {\partial \epsilon_q \over \partial \rho_u }\right )
}_{{\rho_d,\rho_s}} = {\partial \epsilon_q\over \partial \rho_q}
-{1+\delta_{ud}\over \rho_q}\ {\partial \epsilon_q\over \partial \delta_{ud}}
-{1+\delta_{us}\over \rho_q}\ {\partial \epsilon_q\over \partial \delta_{us}},
\nonumber \\
\mu_d &=& {\left ( {\partial \epsilon_q \over \partial \rho_d }\right )
}_{{\rho_u,\rho_s}} = {\partial \epsilon_q\over \partial \rho_q}
+{1-\delta_{ud}\over \rho_q}\ {\partial \epsilon_q\over \partial \delta_{ud}}
-{\delta_{us}\over \rho_q}\ {\partial \epsilon_q\over \partial \delta_{us}},
\nonumber \\
\mu_s &=& {\left ( {\partial \epsilon_q \over \partial \rho_s }\right )
}_{{\rho_u,\rho_d}} = {\partial \epsilon_q\over \partial \rho_q}
-{\delta_{ud}\over \rho_q}\ {\partial \epsilon_q\over \partial \delta_{ud}}
+{1-\delta_{us}\over \rho_q}\ {\partial \epsilon_q\over \partial \delta_{us}},
\label{mui}
\end{eqnarray}
where $\epsilon_q = \sum_{\tau} \epsilon_{\tau }$
 is the total quark energy density
; $\tau=u,d,s$.
The various energy derivatives occuring in Eq. (\ref{mui}) are given by
\begin{eqnarray}
{\left ( {\partial \epsilon_q \over \partial \rho_q } \right ) }_{
\delta_{ud},\delta_{us}} &=& \mu_q =
{1\over 3}\ \sum_{\tau} \lambda_{\tau} f_{\tau} + \ \sum_{\tau} U_{\tau}
, \nonumber \\
{\left ( {\partial \epsilon_q \over \partial \delta_{ud}} \right ) }_{
\rho_q,\delta_{us}} &=&
{\rho_q \over 3}\ \sum_{\tau} \lambda_{\tau}
{\left ( {\partial f_{\tau} \over \partial \delta_{ud} } \right ) }_{
\delta_{us}}, \nonumber \\
{\left ( {\partial \epsilon_q \over \partial \delta_{us}} \right ) }_{
\rho_q,\delta_{ud}} &=&
{\rho_q \over 3}\ \sum_{\tau} \lambda_{\tau}
{\left ( {\partial f_{\tau} \over \partial \delta_{us} } \right ) }_{
\delta_{ud}},
\end{eqnarray}
where
$f_{\tau} = \rho_{\tau}/\rho_b$,
\begin{equation}
U_{\tau}={3\over 2\pi^2 } {\left ( {m_{\tau}c\over \hbar }\right )}^3 \
{\partial m_{\tau}\over \partial \rho_q }
\left [ x_{\tau}\sqrt{1+x_{\tau}^2}-ln(x_{\tau}+\sqrt{1+x_{\tau}^2}
\right ] ,
\label{utau}
\end{equation}
 and $\lambda_{\tau}=m_{\tau}c^2 \sqrt{1+x_{\tau}^2 }$
is the chemical potential of the quarks with density independent
masses which is also the usual bag model result.
The chemical potentials corresponding to each flavour (\ref{mui}) can
also be expressed as follows:
\begin{equation}
\mu_{\tau} = \lambda_{\tau} + \sum_{\tau^{\prime}} U_{\tau^{\prime}}.
\label{mutau}
\end{equation}
Then using Eq. (\ref{mui}), the $\beta-$equilibrium conditions 
(\ref{qbeta}) can
be expressed in terms of the asymmetry parameters as
\begin{eqnarray}
{\partial \epsilon_q\over \partial \delta_{ud}}
-{\partial \epsilon_q\over \partial \delta_{us}} &=& 0,
\nonumber \\
{2 \over \rho_q}\ {\partial \epsilon_q\over \partial \delta_{ud}}
+{1 \over \rho_q}\ {\partial \epsilon_q\over \partial \delta_{us}}
&=& \mu_e,
\label{qbeta1}
\end{eqnarray}
where $\mu_e = {\sqrt {p_{F,e}^2c^2+m_e^2c^4}}$.
>From Eq. (\ref{mutau}), the same can be reexpressed as
\begin{equation}
\lambda_d-\lambda_s=0 \quad {\rm and} \quad \lambda_d-\lambda_u=\mu_e.
\end{equation}
Thus, for a given baryon density $\rho_b = \rho_q/3$, the three quantities
$\rho_e$, $\delta_{ud}$ and $\delta_{us}$
are fixed by the Eqs. (\ref{qcharge}) and (\ref{qbeta1}).
The total energy density of quark matter
$\epsilon_{\rm QM}$ and the pressure $P_{\rm QM}$
for a given $\rho_b$ are then given as
\begin{eqnarray}
\epsilon_{\rm QM} &=& \sum_{\tau}
\epsilon_{\tau} (\rho_q, \delta_{ud},\delta_{us})
+\epsilon_{L} (\rho_q, \delta_{ud},\delta_{us}) , \nonumber \\
P_{\rm QM} &=& \rho_q \ {\partial \epsilon_{\rm QM}\over \partial \rho_q} -
\epsilon_{\rm QM}.
\label{effmeos}
\end{eqnarray}
Eq. (\ref{effmeos}) defines the 
equation of state of charge neutral, $\beta-$
equilibrated quark matter.

In order to make a comparative study, 
we also consider the bag model picture.
The quark matter equation of state within the bag model
can be determined from
Eq. (\ref{effmeos}) itself by setting $C=0$, and adding a bag energy
density $B$ to the total energy density $\epsilon_{\rm QM}$.
One then gets,
\begin{eqnarray}
\epsilon_{\rm bag} &=& \epsilon_{\rm QM}[C=0] + B, \nonumber \\
P_{\rm bag} &=& \rho_q \ {\partial \epsilon_{\rm bag}\over
\partial \rho_q } - \epsilon_{\rm bag}.
\label{bageos}
\end{eqnarray}
Before, we discuss the results obtained within these two models, we give
below the equations of state considered for nuclear matter.

\section{Nuclear matter equation of state }

\subsection{A simple parametrisation }

For the comparative study of properties of hybrid stars
within the bag and effective mass models, we choose a simple
parametrisation of nuclear matter EOS as given in 
Ref. \cite{heisel}.
The total energy density
$\epsilon_n$ for a system of neutrons and protons is taken to be
of the form,
\begin{equation}
\epsilon_n = \rho_b \left [ M_n + E_{\rm comp}+E_{\rm sym} \right ]
,\end{equation}
where $M_n$ is the nucleon mass.
The compressional energy is given by
$E_{\rm comp}=(K_o/18){(\rho_b/\rho_o-1)}^2$
and the symmetry energy  is
 $E_{\rm sym }= S_o \ (\rho_b/\rho_o)$ ${(1-2x_p)}^2$,
where $x_p$ is the proton fraction.
Values for the nuclear matter incompressibility $K_o$, symmetry
energy coefficient
$S_o$ and the normal nuclear matter density $\rho_o$ are
taken to be 250 MeV, 30 MeV and 0.16$fm^{-3}$ 
respectively \cite{heisel}.

The optimum value of $x_p$ for a given $\rho_b$ is determined
by the $\beta-$ equilibrium condition and the charge neutrality
condition:
\begin{equation}
\mu_n=\mu_p+\mu_e \quad {\rm and} \quad \rho_e = x_p \rho_b ,
\label{nmbeta}
\end{equation}
where $\mu_n$ and $\mu_p$ are the neutron and proton chemical
potentials respectively.
The total energy density $\epsilon_{\rm NM}$ and pressure $P_{\rm NM}$
corresponding to nuclear matter for a given baryon density $\rho_b$
is then given as,
\begin{eqnarray}
\epsilon_{\rm NM} &=& \epsilon_n (\rho_b,x_p)+\epsilon_L(\rho_b,x_p),
\nonumber \\
P_{\rm NM} &=& \rho_b {\partial \epsilon_{\rm NM}\over \partial \rho_b}
-\epsilon_{\rm NM}.
\label{nmeos1}
\end{eqnarray}
The above equation defines the equation of state of the $\beta-$
equilibrated, charge neutral nuclear matter.

\subsection{Based on a realistic interaction with explicit
spin degrees of freedom }

The phenomenological momentum and 
density dependent finite range interaction
employed here to obtain 
the equation of state is a modified version of
Seyler-Blanchard interaction \cite{db}.
To treat spin-polarised isospin asymmetric
nuclear matter, 
the interaction has been generalised to include explicitly
the spin-isospin dependent channel.
The interaction between two nucleons with separation $ r $
and relative momentum $ p $ is given by,
\begin{equation}
v_{eff} (r,p,\rho ) = -C_{\tau s} \left [ 1-{p^2\over b^2}-
d^2{\left ( \rho_1(r_1)+\rho_2(r_2) 
\right )}^{\gamma} \right ]{e^{-r/a}\over {r/a}}
\ ,
\label{msb}
\end{equation}
where $ a $ is the range and $ b $
defines the strength of the repulsion in the
momentum dependence of the interaction.
The parameters $ d $ and $ \gamma $ are measures of the strength of the
density dependence, and
$\rho_1$ and $\rho_2$ are the densities at the sites
of the two interacting nucleons. The subscripts $\tau $ and $ s $ in the
strength parameter $C_{\tau s }$
refer to the likeness $ l $ and the unlikeness $ u $ in the isotopic spin
and spin of the two nucleons respectively; for example, $C_{ll}$ refers to
interactions between two neutrons or protons with parallel spins, $C_{lu}$
refers to that between neutrons or protons with opposite spin {\it etc}.
The energy per nucleon $E/A$ and the pressure $P$
in the mean-field approximation can then be worked out \cite{db} as
\begin{equation}
E/A={1\over \rho}
\sum_{\tau s} \rho_{\tau s} \left [ T {J_{3/2} (\eta_{\tau s})
\over J_{1/2} (\eta_{\tau s} ) } (1-m^*_{\tau s} V_{\tau s}^1 ) +
{1\over 2} V_{\tau s}^0 \right ]
\ ,
\label{msbe}
\end{equation}
\begin{equation}
P=\sum_{\tau s} \rho_{\tau s} \left [ {2\over 3}T {J_{3/2} (\eta_{\tau s})
\over J_{1/2} (\eta_{\tau s} ) } + V_{\tau s}^0 +
{1\over 2}b^2 \left (1-d^2(2\rho )^{\gamma} \right ) V_{\tau s}^1 
+ V_{\tau s}^2
\right ] \ .
\label{msbp}
\end{equation}
Here, $J_k (\eta )$ are the Fermi integrals,
$V_{\tau s}^0$ and $V_{\tau s}^2$
are the single-particle and the
rearrangement potentials, $V_{\tau s}^1$ is the coefficient of the quadratic
momentum dependent term in the potential
and defines the effective mass
$m^*_{\tau s}$, $T$ the temperature, and $\eta$
is the fugacity given by
$\eta_{\tau s} = (\mu_{\tau s} - V_{\tau s}^0 - V_{\tau s }^2 )/T $.
For the unpolarised nuclear matter (UNM),
the expressions for $V_{\tau s}^0$ {\it etc} are given in
 ref. \cite{db}. It is straightforward to extend these to the case of
polarised nuclear matter (PNM) and 
are given in detail in Ref. \cite{Um97}.

One usually defines the neutron and proton spin excess parameters ( spin
asymmetry ) as
\begin{eqnarray}
\alpha_{n} &=&  ({\rho_{n \uparrow}-\rho_{n \downarrow}})/\rho \ , \nonumber \\
\alpha_{p} &=&  ({\rho_{p \uparrow}-\rho_{p \downarrow}})/\rho \ ,
\end{eqnarray}
where
\begin{equation}
\rho = \rho_n +\rho_p = (\rho_{n\uparrow}+\rho_{n\downarrow}) +
			(\rho_{p\uparrow}+\rho_{p\downarrow})\ ,
\end{equation}
is the number density. We then define the proton fraction as
$x=\rho_p/\rho$. It is related to the isospin asymmetry parameter $X$
as,
\begin{equation}
X=(1-2x) = (\rho_n - \rho_p )/\rho \ .
\end{equation}
We also define the spin excess parameter as $Y=\alpha_n+\alpha_p$ and the
spin-isospin excess parameter as $Z=\alpha_n - \alpha_p $.
One can then express the energy per nucleon $E/A$ of the NM
at zero temperature as
\begin{equation}
E/A =E_{V}+E_{X} X^2 +E_{Y} Y^2 +E_{Z} Z^2
\ ,
\end{equation}
where terms higher than those
quadratic in $X$, $Y$ and $Z$ are neglected. Here
$E_V$ is the volume energy of the symmetric nuclear matter, taken as
$-16.1$ MeV
and $E_X$ is the usual symmetry (isospin) energy, taken to be $33.4$ MeV. The
quantities $E_Y$ and $E_Z$ are the spin and the spin-isospin symmetry
energies of the NM respectively. 
We take \cite{ph,arh,Um97}
$E_X = 33.4$ MeV,  $E_Y = 15$ MeV and $E_Z=36.5$ MeV.
Values of these coefficients are uncertain to an extent.

The generalised hydrodynamical model of Uberall \cite{ub},
which gives ${( E_Z/E_X )}^{1/2} \simeq 1.1 $ fixes $E_Z$ for a given
value of $E_X$.
The value of $E_Y$ is relatively more uncertain, and it was found 
in our earlier study \cite{Um97} on neutron stars that standard 
neutron star properties are best explained with $E_Y=15$ MeV and 
hence we use here the same set of parameters.
The values of the parameters 
$C_{\tau s}$, $a$, $b$, $d$ and $\gamma$ are then determined by
reproducing $E_V$, $E_X$, $E_Y$, $E_Z$, the saturation density of
normal nuclear matter $(\rho_0 = 0.1533\ fm^{-3}) $, the surface energy
coefficient ($a_S = 18.0$ MeV ), the energy dependence of the real
part of the nucleon-nucleus optical potential and
the breathing-mode energies \cite{mmm}.
The parameters of the interaction so determined are listed below:
\begin{eqnarray}
C_{ll} = -305.2\ {\rm MeV} &\quad& a=0.625\ {\rm fm} \nonumber \\
C_{lu} = \ 902.2\ {\rm MeV} &\quad& b=927.5\ {\rm MeV/c} \nonumber \\
C_{ul} = \ 979.4\ {\rm MeV}  &\quad& d=0.879\ {\rm fm^{3n/2}} \nonumber \\
C_{uu} = \ 776.2\ {\rm MeV}  &\quad & \gamma=1/6 .\nonumber
\end{eqnarray}
With the above value of the parameter $\gamma$, the
incompressibility of symmetric nuclear matter is $K=240 $ MeV.

It may be said here that the above interaction reproduces
quite well the ground state binding energies, root mean square charge radii,
charge distributions and giant monopole resonance energies for a host of
even-even nuclei ranging from $^{16}O$ to very heavy systems.
We have also seen that
for symmetric nuclear matter, our results agree extremely
well with those calculated in a variational approach by
Friedmann and Pandharipande(FP)\cite{fp} with $v_{14}+TNI $ interaction in
the density range ${1\over 2}\rho_0 \le \rho \le 2\rho_0 $.
But, for unpolarised pure neutron matter, the energies calculated \cite{Um97}
with our interaction are somewhat higher compared to the FP energies,
particularly at higher densities. 
However, our results \cite{Um97} 
are very similar to those obtained from the recent
sophisticated calculation of Wiringa et al \cite{rbw} with UV14+UVII
interaction, and hence the present interaction can be extrapolated
with some confidence to neutron matter at high densities.
Then, using this interaction it was found \cite{Um97} that neutron
matter undergoes a ferromagnetic transition at a density 
$\rho \sim 4\rho_0$. \hfill
Before, we investigate the effect of such a 
ferromagnetic transition on neutron star properties, certain general aspects
of SQM as obtained within the effective mass model of confinement
is discussed below.

\section {Flavour symmetric strange quark matter }

Having clearly specified,  in the above two sections, 
the various equations of state 
considered for quark matter and nuclear matter, we are now 
in a position to make a detailed study of hybrid stars.
To do so, firstly, we would like to demonstrate 
the consistency of our present QM calculation done
within the effective mass model, and as well make a
 preliminary study of the general aspects
of the strange quark matter equation of state 
given by Eq. (\ref{effmeos}).
For this purpose, we take the current quark masses of $u$, $d$ and $s$
to be zero. As the parameters $n$ and $C$ (defined in 
Eq. (\ref{effm}) by the density dependence of the quark
masses) are flavour independent,
the quark masses are the same and hence, the matter is flavour
symmetric, {\it i.e.} $f_u=f_d=f_s=1$.

The mass parameters $n$ and $C$ are chosen so that at the density
where pressure $P=0$, ${(M/A)}_{P=0} < 930$ MeV. Here $M/A$ is the
total mass per baryon of the flavour symmetric SQM. In Table 1, the
values of ${(M/A)}_{P=0}$ so-obtained are shown for several values
of $n$ and $C$. The main result of this preliminary investigation
is shown in Fig. 1, where we have plotted $M/A$ and $P$ as a function
of the baryon density $\rho_b$. Values of $n$ and $C$ are taken to be
$2/3$ and $85\ MeV\ fm^{-2}$.
As expected, the minimum of $M/A$ and $P=0$ occurs at the same
density point for all values of $n$ and $C$ considered here.
It may be noted that in the study in Ref. \cite{ben},
the above mentioned density points did not coincide;
this could be due to the noninclusion of the rearrangement
term in the single particle energy.
The role of this rearrangement energy in quark chemical potential
is illustrated in Fig. 2, where the ratio 
$\mu_d/\lambda_d$ is displayed as a function
of baryon density $\rho_b$. 
The chemical potentials $\lambda_d$ and $\mu_d$ are
as defined in Eq. (\ref{mutau}). 
The deviation from unity in the case of the
effective mass model is due to the rearrangement term.
The effect of this term is dominant in the low density regime. As the
density $\rho_b$ increases, $\mu_d/\lambda_d \longrightarrow 1$,
the bag model result; this is due to the fact that
as $\rho_b \longrightarrow \infty$,
$m_{\tau} \longrightarrow m_{\tau}^0$, and hence the
rearrangement term $U_{\tau } \longrightarrow 0$ as
can be seen from Eq. (\ref{utau}).
Similar curves for other flavours are not shown as 
$\mu_s/\lambda_s = \mu_d/\lambda_d= \mu_u/\lambda_u $.

We have also studied the dependence of the properties of strange quark
star on the mass parameters $n$ and $C$. The total mass and the radius
of the star are obtained by solving the general relativistic
Tolman-Oppenheimer-Volkoff (TOV) equation \cite{shapiro},
\begin{equation}
{dP(r)\over dr} = -{G\over c^4} \
{\left [\epsilon (r)+P(r)\right ] \left [ m(r)c^2 + 4\pi r^3 P(r) \right ]
\over { r^2 \left [ 1-{2Gm(r)\over rc^2} \right ] }},
\label{tov}
\end{equation}
where,
\begin{equation}
m(r)c^2 = \int_0^r \epsilon (r^{\prime}) d^3r^{\prime }.
\end{equation}
The quantities $\epsilon (r)$ and $P(r)$ are the energy density and pressure
at a radial distance $r$ from the centre of the star,
and are given by the equation of state (\ref{effmeos}).
The mass of the star contained within a distance $r$
is given by $m(r)$.
The size of the star is determined by the boundary condition $P(R)=0$
and the total mass $M$ of the  star integrated upto the surface $R$
is given by $M=m(R)$. The single integration constant needed to solve
the TOV equation is $P_c$, the pressure at the center of the star
calculated at a given central density $\rho_c$.
The values of the total mass $M_{\rm max}$ and the radius $R$ of the
star corresponding to the maximum mass configuration are given in
Table 1. The central density $\rho_c$ in units of $\rho_o$ is also
given. In order to understand the behaviour of the various structural
properties of SQS with respect to the change in the mass parameters
$n$ and $C$, we have calculated the incompressibility $K_q=(dP/d\rho_q)$
of the star matter, which is displayed in Fig. 3 as a function of
$\rho_b$ for three sets of $(n,C)$. It is found that with increase in $n$
 (keeping $C$ fixed), the EOS becomes stiffer. With $n$ fixed, if $C$ is
increased, the EOS becomes effectively a little stiffer(averaged over
the density of the star) though at low densities, it is somewhat softer.
>From Table 1, we note that as the EOS becomes stiffer, ${(M/A)}_{P=0}$
and $\rho_c$ increases whereas $R$ and $M_{\rm max}$ decreases.
These results corroborate the findings by Haensel et al \cite{haensel}
in the bag model. We expect these features to be valid even when the
$\beta-$equilibrium conditions are considered, since it is in general
found \cite{haensel}
that $\beta-$equilibrated SQM is nearly flavour symmetric and
the electron density is negligibly small.

Finally, taking into account both $\beta-$equilibrium and charge 
neutrality conditions given by equations (\ref{qbeta}) and
 (\ref{qcharge}) respectively,  we determined the mass-radius 
relationship of the strange quark star by solving the TOV equation 
in both the models described by Eqs. (\ref{effmeos}) and (\ref{bageos}).
Results so obtained are displayed in Fig. 4 and it can be seen that
results obtained within the two models of quark confinement are
quite similar.
It may be stressed here that the values of
$M_{\rm max}$ and $R$ are dependent on the model parameters.
It is found that with suitable choice of the parameters,
it is possible to obtain the maximum mass and the radius
of the star well within the acceptable limits.
For the sake of comparison, we have shown the results
obtained for a neutron star in the same figure.
The structure of the neutron star is determined using
a composite EOS, {\it i.e.}
Feymann-Metropolis-Teller \cite{fmt},
Baym-Pethick-Sutherland \cite{bps},
Baym-Bethe-Pethick \cite{bbp} and that given by 
Eqs. (\ref{msbe}-\ref{msbp})
with progressively increasing densities. It is already
known from previous studies \cite{haensel} that curvature of
the $M-R$ curves in the low mass region as obtained
from the bag model for SQS is opposite to that of
the neutron star. This difference is also borne out
in the present calculation using effective quark
masses.
Thus, our present calculation of quark matter EOS
within effective mass model is consistent with the
general findings in literature.

\section{ Hybrid stars : A comparative study of two models
of quark confinement } 

Strange quark matter may not be the absolute ground state of
hadronic matter. If this is so, then it is possible that
neutron stars are not made up of only quark matter, but have
a quark core enveloped by nuclear matter, usually referred to
as hybrid stars. Since the quark-hadron transition is normally taken to
be of first order, there exists a mixed phase consisting of
droplets of quark matter and of nuclear matter in thermodynamic
equilibrium. As this occurs at a unique pressure, such a phase
cannot exist inside the hybrid star. Taking into account the
recent viewpoint of Glendenning \cite{glen} as already
mentioned in the introduction, the existence of this droplet
phase inside the hybrid star cannot however be ruled out.
In this section, we investigate the structure of this droplet
phase and its observable consequences on the structural
properties of the neutron star using the two models of
quark confinement, {\it viz} bag and effective mass models.
We thereby perform calculations similar to the
earlier studies \cite{ben,chakra}, but
with the correct treatment of quark chemical potentials. 
Moreover, we make use of the simple parametrisation of nuclear
matter EOS given by Eq. (\ref{nmeos1}) in order to be able to
directly compare our results to those 
obtained in earlier studies \cite{heisel,chakra}.

\subsection{Equation of state for droplet phase}

We consider the droplet phase to be present in a uniform
background \cite{heisel} of electron gas. The system as a whole is
charge neutral, and each phase of matter is subjected
to appropriate $\beta-$equilibrium conditions.

For the nuclear matter phase, we consider the EOS as given
by Eq. (\ref{nmeos1}). 
The appropriate $\beta-$ equilibrium condition
is $\mu_n=\mu_p+\mu_e$. The total charge density $\rho_z^N$
pertaining to the NM phase is given by
$\rho_z^N=\rho_p-\rho_e=x_p \rho_b^N-\rho_e$, where $\rho_p$
is the proton density and $\rho_b^N$ is the total baryon
density in the NM phase.
For given values of $\rho_b^N$ and
electron density $\rho_e$, the proton
fraction $x_p$ is determined from the $\beta-$ equilibrium
condition.
In the case of quark matter phase, the equations of state
are given by Eqs. (\ref{effmeos}) and (\ref{bageos}).
The total charge density $\rho_z^Q$ in this phase is
$\rho_z^Q=(2\rho_u-\rho_d-\rho_s)/3-\rho_e $, where the flavour
densities are as defined in Eq. (\ref{rhoi}).
For given values of
$\rho_b^Q(=\rho_q/3=(\rho_u+\rho_d+\rho_s)/3)$ and
$\rho_e$, the asymmetry parameters $\delta_{ud}$ and $\delta_{us}$
are determined from the $\beta-$ equilibrium conditions given
by Eq. (\ref{qbeta1}).

The two conserved charges of this uniform droplet phase are the
total baryon number and the total charge.
The corresponding baryon density $\rho_b$ and
charge density $\rho_z$
are expressed in terms of the volume fraction $\chi$ as:
\begin{eqnarray}
\rho_z &=& \chi \rho_z^Q + (1-\chi ) \rho_z^N, 
\label{rhoz}
\\
\rho_b &=& \chi \rho_b^Q + (1-\chi ) \rho_b^N.
\label{rhob}
\end{eqnarray}
Global charge neutrality requires that $\rho_z=0$.
Thus for given values of $\rho_b$, $\chi$ and $\rho_b^N$, the
electron density $\rho_e$ is fixed by charge neutrality condition
given by Eq. (\ref{rhoz}) and $\rho_b^Q$ is determined from 
 Eq. (\ref{rhob}).
To arrive at the EOS of the droplet phase at a given density $\rho_b$,
$\chi$ and $\rho_b^N$ are to be determined from mechanical
and chemical equilibria. The chemical equilibrium between the
NM and QM phases requires that $\mu_n=2\mu_d+\mu_u$ and
$\mu_p=\mu_d+2\mu_u$. From mechanical equlibrium, one has
$P_{\rm QM}=P_{\rm NM}$, where the pressures in the two
phases are as defined in Eqs. (\ref{nmeos1}) and 
(\ref{effmeos},\ref{bageos}).
The total energy density $\epsilon_{T}$ of the droplet phase is
given by
\begin{equation}
\epsilon_T = \chi \epsilon_{\rm QM} + (1- \chi ) \epsilon_{\rm NM},
\end{equation}
where $\epsilon_{\rm QM}$  and $\epsilon_{\rm NM}$
are calculated with the constraints of mechanical, chemical and
$\beta-$ equilibrium conditions as well as with the condition of
global charge neutrality.

For the study of the hybrid stars, the parameters in the
effective mass model and the bag model are chosen so
that ${(M/A)}_{P=0} > 930$ MeV. The parameters chosen in
the effective mass model are $n=2/3$, $C=125$ MeV $fm^{-2}$ and
$m_s^0=250$ MeV. In the bag model, we have taken $B^{1/4}=175$
MeV and $m_s^0=250$ MeV.

\subsection{Results and Discussions}

The droplet phase obtained with global charge neutrality
condition (as detailed in Sect. V.A.) is found to be
energetically more favourable than the mixed phase
determined by the Gibbs criteria, where each of the two
phases is separately charge neutral.
This is illustrated, in the effective mass and bag models,
in Figs. 5 and 6 respectively. The dashed line corresponds
to the mixed phase as obtained by the common tangent method.
It is seen that the droplet phase extends
well beyond the mixed phase (dashed line).
This was also noted in earlier studies \cite{glen,heisel}.
In the limits of
low density $(\chi \longrightarrow 0)$ and high density
 $(\chi \longrightarrow 1)$, the droplet phase smoothly
joins with the NM and QM equations of state respectively
as it should.

It may be parenthetically noted that the ``down-turn''
behaviour in the quark matter EOS in the low density region
in Fig. 5 is due to the choice of the parameter $n$.
In the limit $\rho_b \longrightarrow 0$, the quark
energy density $\epsilon_q$ [Eq. (\ref{etau})]  behaves as,
\begin{eqnarray}
\epsilon_q & \sim & {\rho_q^{5/3}\over m_q } + m_q \rho_q,
\nonumber \\
\quad & \sim & \rho_b^{n+5/3} + C\rho_b^{1-n}.
\end{eqnarray}
If $n=1$, in the limit $\rho_b \longrightarrow 0$,
$\epsilon_q \longrightarrow C$ in agreement with the
usual bag model result. For $n=2/3$,
in the limit $\rho_b \longrightarrow 0$,
$\epsilon_q \longrightarrow 0$, and this causes the ``down-turn''
in the quark matter EOS.
We have chosen $n=2/3$ as the
Dirac-scalar potential [Eq. (\ref{effm})] is generally taken to be of
harmonic oscillator type, {\it i.e.} $l=2$.

We now explore the possible consequences
of the presence of a droplet phase on the structure of
the hybrid star. To do this, we solve the TOV equation
using the EOS shown in Figs. 5 and 6. The baryon number
density and mass distributions inside the star,
corresponding to the maximum mass configuration,
determined using the effective mass and bag models are
shown in Figs. 7 and 8 respectively. The hybrid star
structure in the effective mass is characterised by the
central density $\rho_c = 0.83\ fm^{-3}$, the radius
$R=11.8\ km$ and maximum mass $M_{\rm max}= 1.48 M_{\odot}$.
In the bag model we have
$\rho_c = 0.88\ fm^{-3}$, $R=11.6\ km$ and
$M_{\rm max}= 1.52 M_{\odot}$.
One sees from Figs. 7 and 8 that the droplet phase extends in
both the models from $\sim 1\ km$ to $\sim 8\ km$, and stars
contain a core of about $1\ km$ made of only QM and a crust
region of about $4\ km$ made up of only NM. We find that the
droplet phase encompasses most of the volume of the star in
agreement with an earlier study \cite{glen}.
In a recent study \cite{drago} using the color-dielectric
model, it was however noted that the substantial part
of the core of the star contains only QM, and the
droplet occupies roughly $40\% $ of the total volume.
It must be stressed that these finer details are in
general model-dependent. 
Finally, the main conclusion of this section is that 
with the correct treatment of quark chemical potentials, 
we find that the hybrid star properties as obtained within the 
bag and effective mass models are quite similar.
For this reason, for the study on the effect of 
spin polarisation on droplet phase, done in the
following section, we use only the 
standard bag model.

\section{Spin polarised nuclear matter and hybrid stars}
In this section, we explore the properties of the hybrid
stars using the proposed NM equation of state given
by Eqs. (\ref{msbe}) and (\ref{msbp}), which predicts
a ferromagnetic phase transition at high densities.

\subsection{Spin polarisation and droplet phase}

Firstly, to know whether there exists a phase transition 
between the nuclear matter phase described by  
Eqs. (\ref{msbe},\ref{msbp}) and the QM phase characterised
by the bag model EOS, Eq. (\ref{bageos}), we compare the
total energies per baryon obtained in the two phases.
The parameters chosen for this purpose are as given in 
section III.B for NM phase, and for QM, we have taken
$B^{1/4}=180$ MeV and $m_s=250$ MeV.
Further, in the case of NM, the total energy is 
minimised with respect to both the neutron and
proton spin polarisation parameters, $\alpha_n$ 
and $\alpha_p$ respectively. This energy minimised nuclear 
matter calculation
is hereafter referred to as EMNM. In addition, 
we also calculated the total energies for 
unpolarised nuclear matter (UNM), where 
$\alpha_n =0$ and $\alpha_p=0$ at all densities.
Results of these calculations are displayed in
Fig. 9.

One can observe a slight bend in the EMNM curve at 
$\rho \sim 0.7 \ fm^{-3}$. This point indicates
the ferromagnetic phase transition density, 
$\rho_{\rm FM }$, where the optimum value of
$\alpha_n$ sharply increases from zero to unity.
( This feature is clearly illustrated in
Fig. 3 of Ref. \cite{Um97}. )
It is also seen that a polarised nuclear matter (PNM)
phase is energetically favourable over the density
region $\rho_{\rm FM} \le \rho_b \le \rho_{\rm HQ}$, 
where $\rho_{\rm HQ} \simeq 10\rho_0$ is the 
NM-QM phase transition density. It would be interesting
to know whether there exists a
droplet phase consisting of droplets of SQM and PNM.
For this purpose, we made an attempt to solve for the droplet
phase following the same procedure detailed in section V.
It was then found that there exists no solution for 
phase equilibrium between SQM and PNM phases.
In order to understand this result, we do the following
analysis.

Firstly, to show the effect of increasing proton fraction
on $\alpha_n$, we calculated, for given values of $x_p$
and $\rho_b$, the optimum value of $\alpha_n$ by energy
minimisation. 
Values of $\alpha_n$ so determined are plotted in Fig. 10
as a function of density for few values of $x_p$.
( It may be said here that for $x_p \le 0.5$, values 
usually found in the droplet phase, the 
maximum of the optimum value of $\alpha_p$, 
obtained with increase in density, is very nearly zero. )
As $x_p$ increases, the value of $\rho_{\rm FM }$ 
increases and the maximum value of $\alpha_n$, 
given by $(1-x_p)$, decreases.
Therefore, in general, one might say that NM phase with
large value of $x_p$ disfavours spin polarisation. 
On the otherhand, as it was noted \cite{Um97} before, 
as NM gets spin polarised the $\beta-$equilibriated
$x_p$ sharply falls to zero,  which
is also evident from Fig. 11, where we have shown mass 
per baryon versus density for certain cases.
One can see that there is a slight deviation,  
for densities $\rho_b < \rho_{\rm FM }$, between
the EMNM (with $x_p=0$) curve and the EMNM curve 
obtained taking $\beta-$equilibriation into
account, which is essentially due to non-zero, 
though small, values of $x_p$. However, as density
is increased further, the two curves coincide
exactly illustrating the fact that $\beta-$equilibriated
$x_p$ values become zero at $\rho \simeq \rho_{\rm FM}$.
Another interesting aspect is that beyond the ferromagnetic
transition density, the PNM phase is energetically more
favourable than the UNM with large values of $x_p$.
Hence, for densities $\rho_b > \rho_{\rm FM }$, droplets
of UNM with $x_p > 0$ would prefer to become PNM with
$x_p=0$. That is, one expects a transition from a 
state $(x_p > 0, \alpha_n =0)$ to a
state $(x_p = 0, \alpha_n \simeq 1)$.
But, this change of state, in the context of
formation of a DP, is disallowed for the
following reasons.

In contrast to the usual mixed phase defined
by the common tangent method, 
in the case of a droplet phase, 
the electron density is same both in NM and QM 
phases. Restricting to densities 
$\rho_b > \rho_{\rm FM }$, the following three
cases are considered.\\
\noindent {\it CASE \ 1} : Firstly, let us 
consider the simplest situation with $\mu_e=0$
and $N_e=0$, where we have neglected the small
$e^-$ mass. Then, the optimum configuartion for
QM phase gives $\mu_u = \mu_d = \mu_s$ and
$N_u=N_d=N_s$, where we have taken 
$m_u=m_d=m_s$. 
For the NM phase, at $\rho_b > \rho_{\rm FM }$, 
the optimum configuartion gives $x_p=0$ and
$\alpha_n=1$.
Thus, one finds that the two phases are 
separately charge neutral, and hence 
corresponds to the usual mixed phase
with $N_e=0$. \\
\noindent {\it CASE \ 2} : Secondly, let us 
consider as in case 1, $\mu_e=0$ and $N_e=0$.
But now, we insist that $m_s > m_d = m_u$.
Then the optimum configuration for the
QM phase gives $\mu_u = \mu_d = \mu_s$ with
$N_s < N_d=N_u$, and thereby the QM droplets
are now positively charged. For the global
charge neutrality condition to be satisfied, 
the NM droplets should be negatively charged.
However, according to the optimum configuration, 
the NM droplets consists of only spin polarised
neutrons. Therefore, in this case, there exists
no DP solution. \\
\noindent {\it CASE \ 3} : Finally, let us 
consider the general situation, where
$\mu_e \ne 0$ ( and $N_e \ne 0$ ), and
 $m_s > m_d = m_u$.
Now with finite electron density, the 
$\beta-$equilibrium condition 
pertaining to the NM droplets, at a density
$\rho_b$, is given by 
$(1/\rho_b )\ d\epsilon_{ n} /dx_p 
+ \mu_e =0$. 
Since $\mu_e > 0$, the slope factor 
$d\epsilon_{ n} /dx_p $ must be 
negative. From Fig. 12, it can be 
seen that this is satisfied for the
configuration UNM, i.e. $x_p \sim 0.5$
and $\alpha_n =0$.
This then suggests that 
NM droplets with $\rho_b > \rho_{\rm FM}$
in DP may remain unpolarised with $x_p >0$,
which however is not the optimum configuration
for NM phase.
Thus, due to the nature of droplet phase, 
that is, constancy of electron density 
in both phases, and the interplay of 
$x_p$ and $\alpha_n$, one finds that
NM droplets with $\rho_b > \rho_{\rm FM}$
in DP may remain unpolarised with $x_p >0$.
It may be remembered that the usual mixed phase
between SQM and PNM however exists.
In the following subsection, we explore the
possibility of forming a DP with UNM phase and
SQM phase.

\subsection {Properties of hybrid stars}

>From our earlier study on hybrid stars, 
it can be seen 
( from Table 2 of Ref. \cite{Um97} ) that 
for the chosen set of parameters, the central density
of the star is about $6 \rho_0$. Therefore, the 
UNM-SQM phase transition at $\rho \simeq 0.8\ fm^{-3}$
is of more relevance. Following the same procedure
described in section V, we then solved the set of simultaneous
equations corresponding to $\beta-$equilibrium, charge 
neutrality, chemical and mechanical equilibria.
Results obtained for the droplet phase along with the
usual mixed phase are shown in Fig. 13. 
It can be seen that as in other cases, the droplet phase
extends well beyond the mixed phase. 
Interestingly, the droplet phase is energetically 
 favourable even than the EMNM case over the 
region $0.5 \le \rho \le 0.9 $.
Further, in this particular case, only a part of DP
is allowed as EMNM configuration takes over at high
densities.
Thus, one has according to the minimum energy criterion, 
a DP sandwiched between UNM and PNM phases.
We now explore the possible consequences of the 
presence of a DP on the structure of hybrid stars.

In Fig. 14, we show the proton fraction obtained in
the DP using the realistic interaction [Eq. (\ref{msb})], 
and compare with the one obtained within the
simple parametrisation [Eq. (\ref{nmeos1})].
 It can be seen
that the general trend is quite similar in both
the cases. 
Values of $x_p$ as obtained with the realistic 
interaction is somewhat lower than the simple one.
As the bag model parameters are more or less the
same, this could be due to the fact that the
EOS based on realistic interaction is relatively
softer than the simple parametrised EOS.
Nevertheless, values of $x_p$ is high
enough to allow for direct URCA process to happen.
We now solve the TOV equation (\ref{tov}) using 
the EOS shown in Fig. 13, and obtain the structural
properties such as mass and size of the star.
The baryon number density and mass distribution 
inside the star, corresponding to the maximum 
mass configuration, are plotted in Fig. 15.
The maximum mass configuration is characterised
by the central density $\rho_c = 1.02 \ fm^{-3}$, 
the radius $R=11.4\ km$ and maximum mass 
$M_{\rm max} = 1.62 M_{\odot}$. Now, from Fig. 15, 
one observes that there exists a mixed phase 
consisting of droplets of SQM and UNM sandwiched
between a PNM core and UNM crust.
The droplet phase extends from 
$\sim 2\ km$ to $\sim 6\ km$. 
More interestingly, 
there exists also  a core of about 1 $km$  consisting
PNM. Thus, the salient features of this star 
structure are the following.
Firstly, the mass and size of the star obtained in 
the present calculation is well within the 
acceptable limits. 
Secondly, due to the presence of a droplet phase, 
the value of $x_p$ found inside the star allows
direct URCA process to happen, and thereby
have important consequences on the cooling
process of neutron stars.
Finally, due to the presence of a spin polarised
core, we could also, in principle, explain the
surface magnetic field of neutron stars.
Value of the magnetic field at the surface of
the star due to the presence of a ferromagnetic
core can be estimated using the relation 
$H \sim \alpha_n \mu_n N/R^3$, where 
$ N \simeq 1.4 \times 10^{54} $ is the number of
neutrons in the ferromagnetic phase and 
$R \simeq 11.4\ km$ is the star size.
Then using $\mid \mu_n \mid =1.91$ nuclear
magneton, we get $H \sim 10^{13}\ G$.
Thus, the surface magnetic field is somewhat
overestimated compared to the observed 
value $\sim 10^{12}\ G$.
However, by fine-tuning the parameter $E_Y$, 
as $\rho_{\rm FM}$ is sensitive to this parameter, 
one could arrive at the right value.

Before summarising our findings, we would like to
stress that the star's internal structure very 
much depends upon the model parameters chosen.
For example, keeping $E_Y$ fixed at 15 MeV, and
decreasing the bag parameter from 180 MeV to 170 MeV, 
one can see from Fig. 16 that PNM phase is no more
allowed by energy minimum criterion.
On the otherhand, with the bag
parameter fixed at 180 MeV 
 and increasing $E_Y$ from 15 MeV 
to 17 MeV, one finds that, 
though there exists a small density region 
where DP is favoured, 
PNM phase occupies larger phase space than 
the DP as illustrated in Fig. 17. 
Hence, in this case, the surface magnetic 
field will be largely overestimated.
However, within an allowed set of parameters, we 
could, in principle, quite satisfactorily explain
the structural properties, surface magnetic field
as well as allow for the formation of a DP.

\section{Summary}
To summarize, firstly, 
we have constructed a quark matter equation
of state in the effective mass approximation and then
applied it to investigate some properties of strange quark
matter with particular reference to hybrid and strange
quark stars.
In this effective mass model, the quark masses are chosen
to be density dependent and the absolute confinement of
quarks as well as the asymptotic freedom are adequately
taken care of through this density dependence.
With these effective masses, it is seen that the chemical
potential acquires an extra term, the so-called
rearrangement term. Presence of such a term in the
single particle energy is consistent
with the well-known Hugenholtz-Van Hove theorem.
Thus, the present calculation, in contrast to 
earlier studies, correctly treats the quark chemical
potentials in the effective mass model.

Secondly, we make a comparative study of hybrid star
properties as obtained within the effective mass and
bag models. For nuclear matter equation of state,
we first used a simple parametrisation. 
It is found that a well-defined mixed phase 
of droplets of quark and nuclear
matter sandwiched between a pure quark matter core and
a crust of only nuclear matter exists. This mixed phase
is found to occupy most of the volume of the star. 
The main result of this comparative study is that
both the bag and effective mass models yield similar
results.

Thirdly, using a realistic equation of state of nuclear matter
based on a finite range, momentum and density
dependent (FRMDD) interaction, which predicts a
ferromagnetic phase transition at a density
$\rho_b \sim 4\rho_0$, we investigated the 
effect of spin polarisation on the formation 
of a droplet phase. It is found that as nuclear
matter gets polarised, the value of proton fraction 
 $x_p$ drops sharply to zero, or vice versa, as 
$x_p$ increases, the transition density 
$\rho_{\rm FM}$ increases and hence matter with
$x_p > 0$ prefers to remain unpolarised.
Because of this interplay between $\alpha_n$ 
and $x_p$, there exists no mixed phase
consisting of droplets of strange quark matter 
and polarised nuclear matter.

Finally, using the same equation of state based
on FRMDD interaction alongwith the usual bag model, 
we find that a droplet phase consisting of 
strange quark matter and unpolarised nuclear matter
can however exist. This droplet phase is sandwiched
between a core made up of polarised nuclear matter and
a crust containing unpolarised nuclear matter.
This, in principle, allows for a satisfactory explanation
of the structural properties and surface magnetic field
of neutron stars and keeps room for their rapid cooling
due to the direct URCA process. 

\noindent {Acknowledgements\ :}\ 
The authors are grateful to Amand Faessler for his 
valuable comments and suggestions. One of us (V.S.U.), 
gratefully acknowledges the 
Institute of Theoretical Physics of University of
Tuebingen for financial support, where a part of
this work was completed.

\newpage

\newpage
\begin{figure}
\begin{center}
\leavevmode
\epsfxsize = 8cm
\epsffile[40 85 430 710]{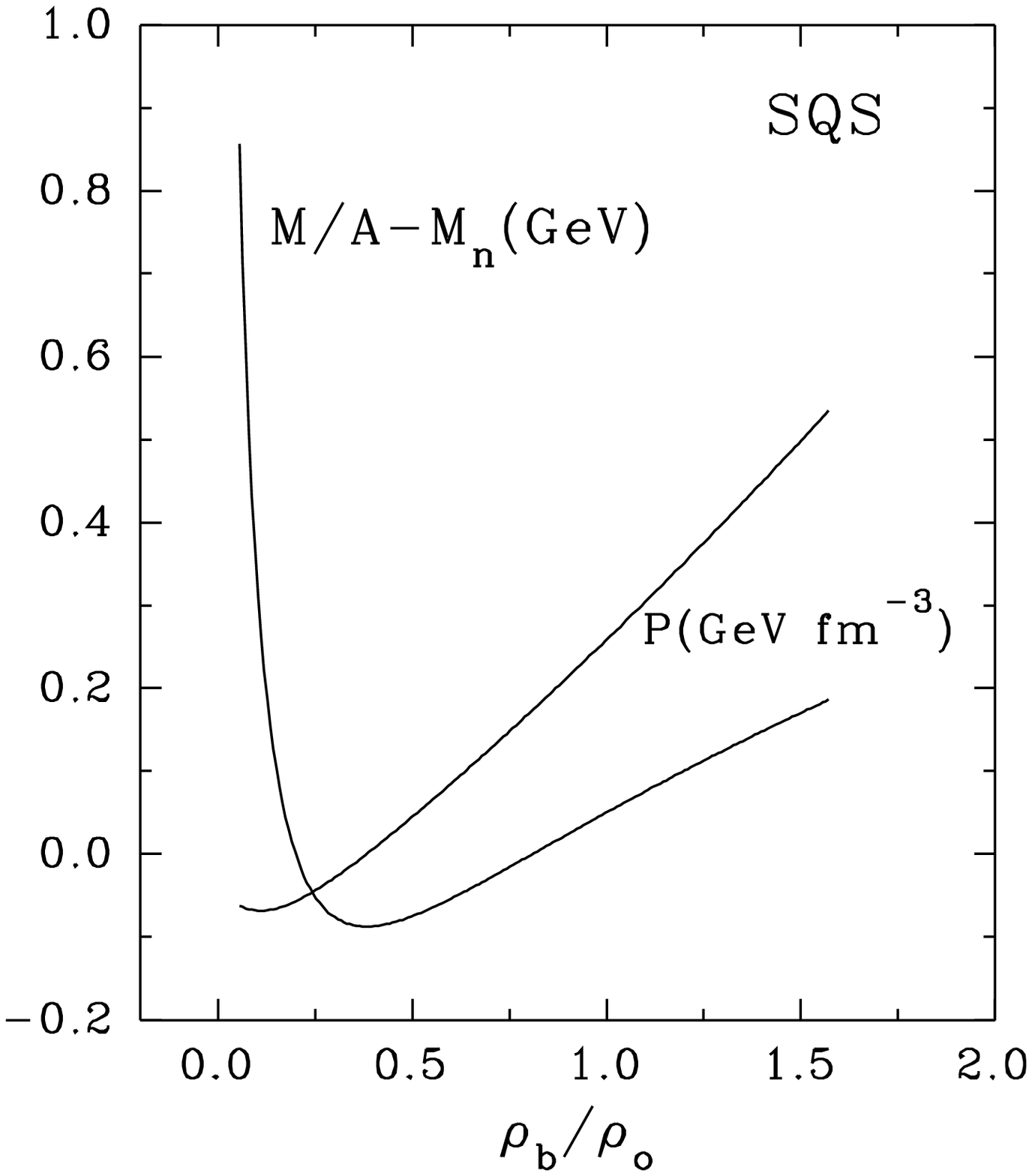}
\end{center}
\caption{
Total mass per baryon $M/A$ and 
total pressure $P$ corresponding
to flavour symmetric strange quark matter is shown
as a function of baryon density 
$\rho_b$ in the effective mass model. $M_n$ is
the nucleon mass and $\rho_o$ is the normal nuclear
matter density.
}
\label{fig1}
\end{figure}
\newpage
\begin{figure}
\begin{center}
\leavevmode
\epsfxsize = 8cm
\epsffile[40 85 430 710]{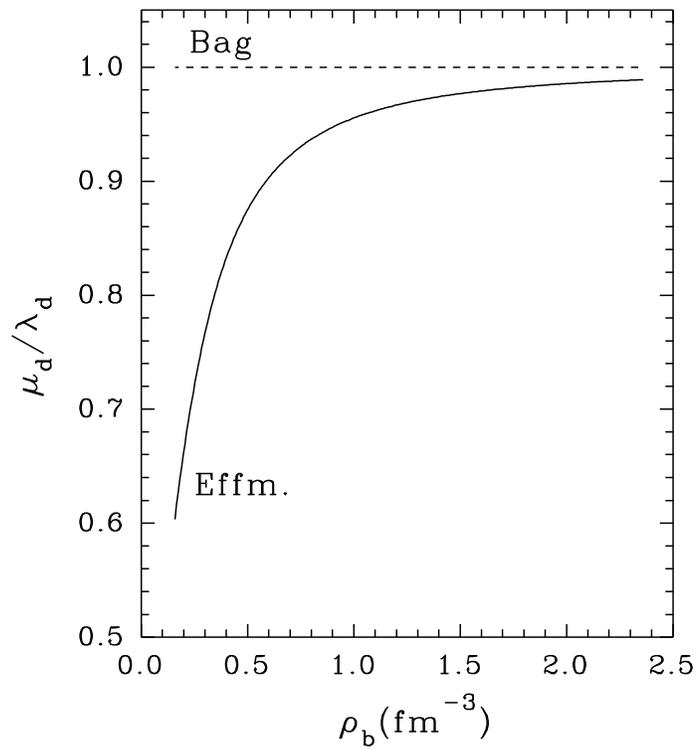}
\end{center}
\caption{
Quark chemical potential [see Eq. (\ref{mutau})]
obtained in the effective mass and bag models are shown
as a function of baryon density $\rho_b$.
}
\label{fig2}
\end{figure}
\newpage
\begin{figure}
\begin{center}
\leavevmode
\epsfxsize = 8cm
\epsffile[40 85 430 710]{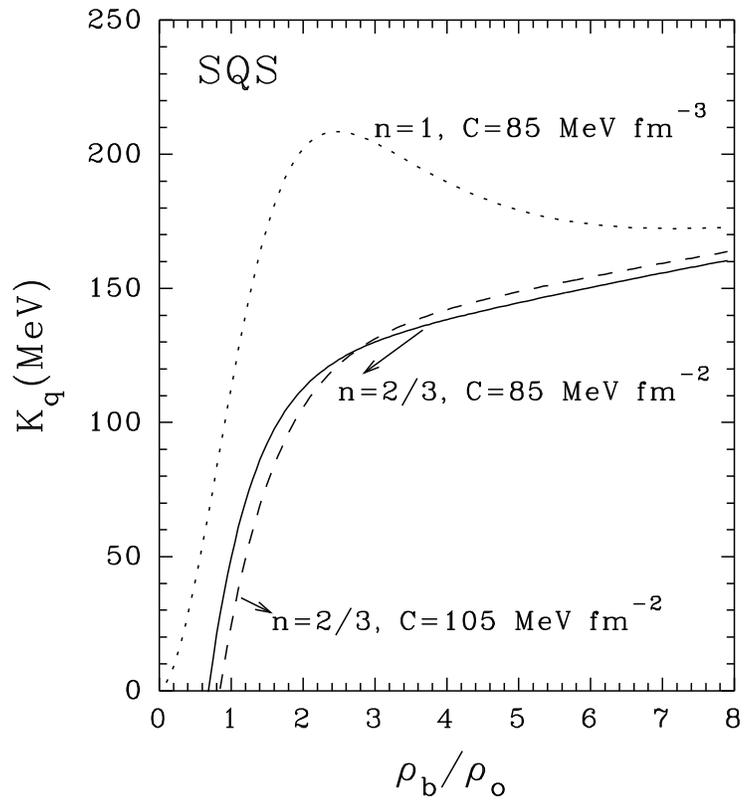}
\end{center}
\caption{
Incompressibility $K_q$ of flavour symmetric strange
quark matter is shown as a function of baryon density
$\rho_b$ in the effective
mass model. Values of the mass parameters $n$ and $C$
chosen are given.
$\rho_o$ is the normal nuclear
matter density.
}
\label{fig3}
\end{figure}
\newpage
\begin{figure}
\begin{center}
\leavevmode
\epsfxsize = 8cm
\epsffile[40 85 430 710]{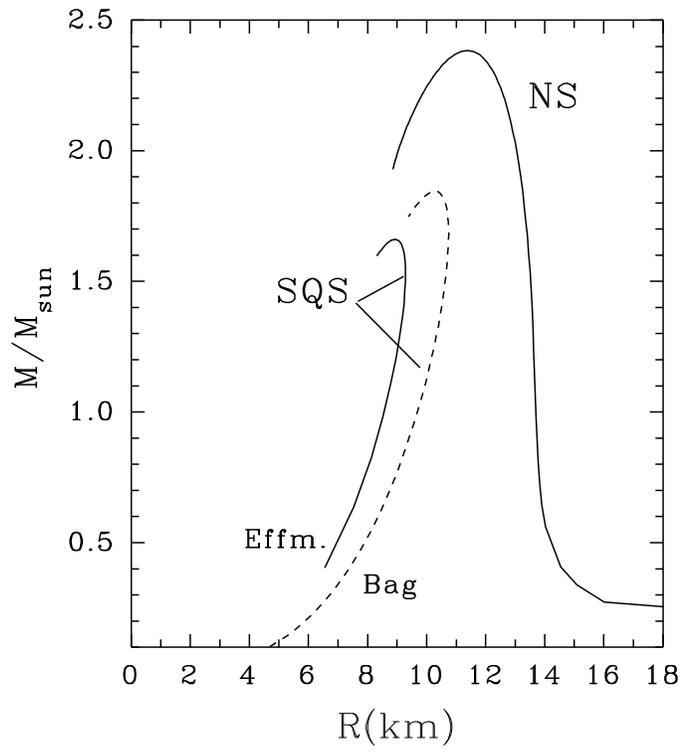}
\end{center}
\caption{
Mass-radius relationships of strange quark stars (SQS) as obtained
in effective mass and bag models are compared with a neutron
star (NS) one.
}
\label{fig4}
\end{figure}
\newpage
\begin{figure}
\begin{center}
\leavevmode
\epsfxsize = 8cm
\epsffile[40 85 430 710]{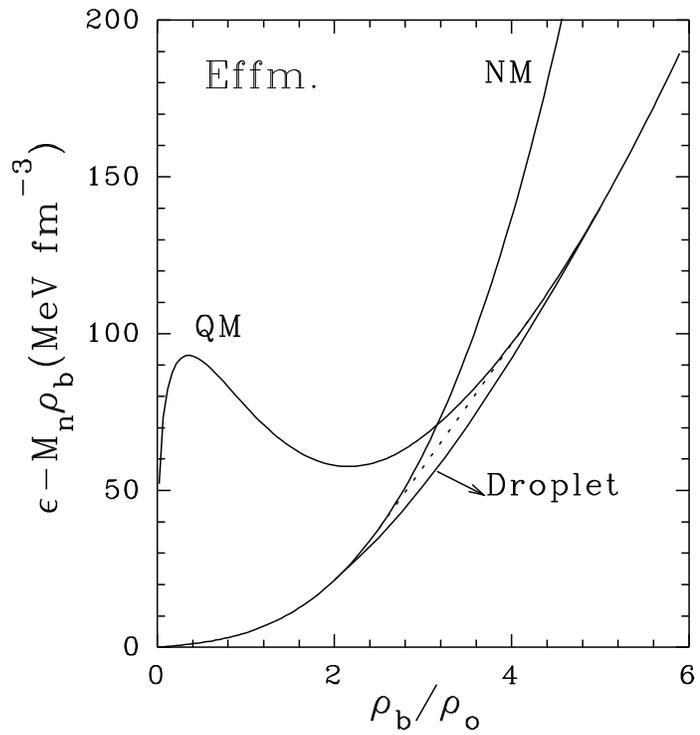}
\end{center}
\caption{
Total energy densities of nuclear matter (NM), quark matter (QM) and
droplet phase are shown
as a function of baryon density $\rho_b$ in the effective mass model.
The dashed line represents the mixed phase
obtained by common tangent
method.
$M_n$ is the nucleon mass.
}
\label{fig5}
\end{figure}
\newpage
\begin{figure}
\begin{center}
\leavevmode
\epsfxsize = 8cm
\epsffile[40 85 430 710]{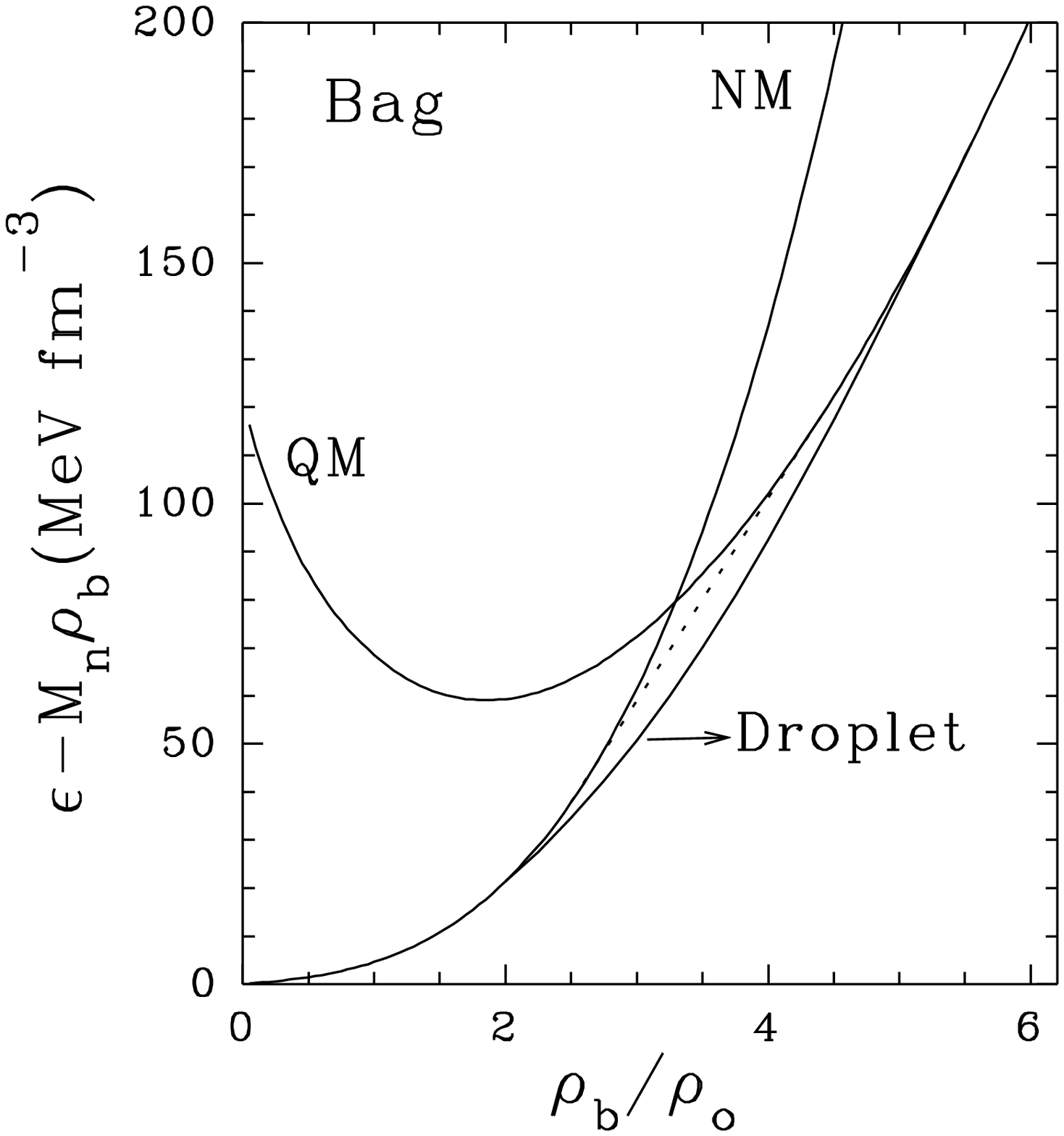}
\end{center}
\caption{
Same as in Fig. 5, but for the bag model.
}
\label{fig6}
\end{figure}
\newpage
\begin{figure}
\begin{center}
\leavevmode
\epsfxsize = 8cm
\epsffile[40 85 430 710]{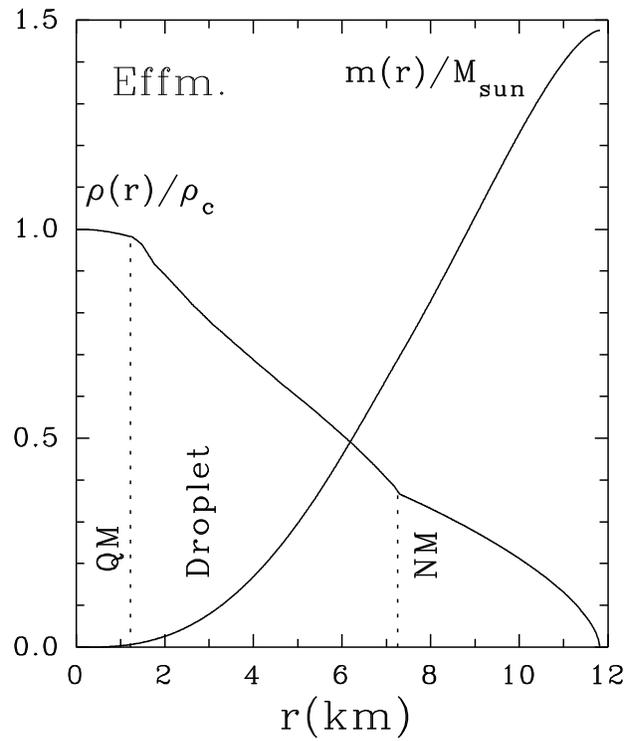}
\end{center}
\caption{
Baryon density $[\rho (r)]$ and mass $[m(r)]$ distributions
corresponding to the maximum mass configuration
are shown
as a function of the radial distance $r$ in the effective mass model.
$M_{\rm sun}(\equiv M_{\odot })$ is the solar mass and
$\rho_c$ is the baryon density at the core $(r=0)$ of
the star.
}
\label{fig7}
\end{figure}
\newpage
\begin{figure}
\begin{center}
\leavevmode
\epsfxsize = 8cm
\epsffile[40 85 430 710]{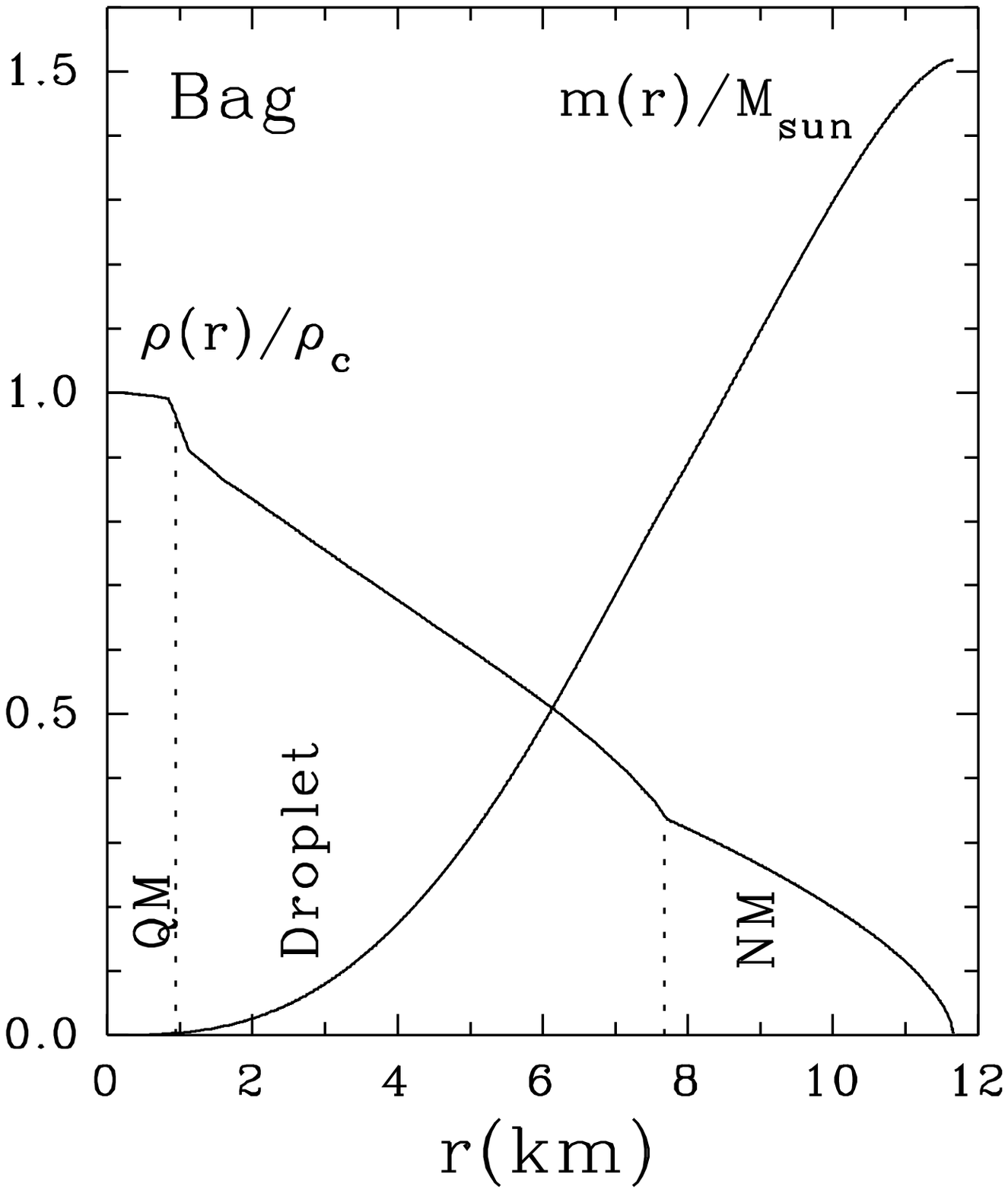}
\end{center}
\caption{
Same as in Fig. 7, but for the bag model.
}
\label{fig8}
\end{figure}
\newpage
\begin{figure}
\begin{center}
\leavevmode
\epsfxsize = 8cm
\epsffile[40 85 430 710]{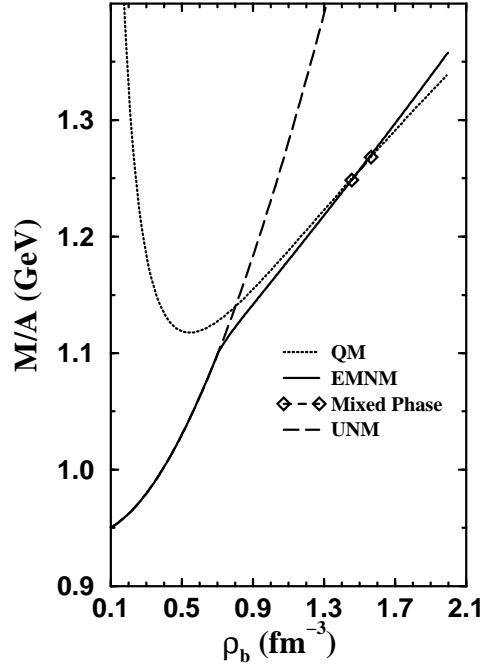}
\end{center}
\caption{
Total mass energies per baryon 
of quark matter (QM), nuclear matter minimised 
with respect to spin polarisation 
parameters (EMNM), unpolarised nuclear
matter (UNM) and
mixed phase are shown
as a function of baryon density $\rho_b$.
Parameters chosen are : $B^{1/4}=180\ MeV$, 
$m_s=250\ MeV$ and $E_Y = 15\ MeV$.
}
\label{fig9}
\end{figure}
\newpage
\begin{figure}
\begin{center}
\leavevmode
\epsfxsize = 8cm
\epsffile[40 85 430 710]{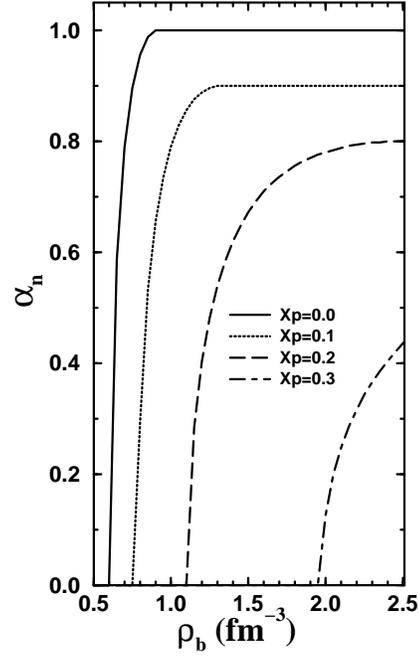}
\end{center}
\caption{
The value of spin polarisation parameter $\alpha_n$
obtained from minimisation of energy for particular
values of $x_p$.
}
\label{fig10}
\end{figure}
\newpage
\begin{figure}
\begin{center}
\leavevmode
\epsfxsize = 8cm
\epsffile[40 85 430 710]{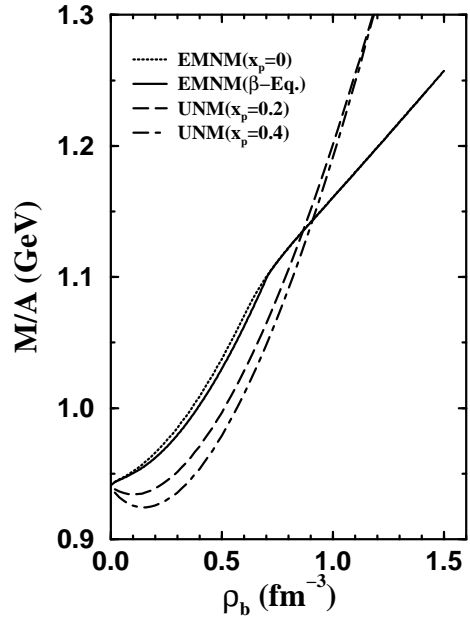}
\end{center}
\caption{
Mass energies per baryon obtained for pure neutron
matter $(x_p=0)$ and $\beta-$equilibriated 
nuclear matter (solid line)
considering energy minimisation with respect to
spin polarisation parameters are shown.
Results obtained for unpolarised nuclear matter (UNM)
with particular values of proton fraction $x_p$ 
are also plotted.
}
\label{fig11}
\end{figure}
\newpage
\begin{figure}
\begin{center}
\leavevmode
\epsfxsize = 8cm
\epsffile[40 85 430 710]{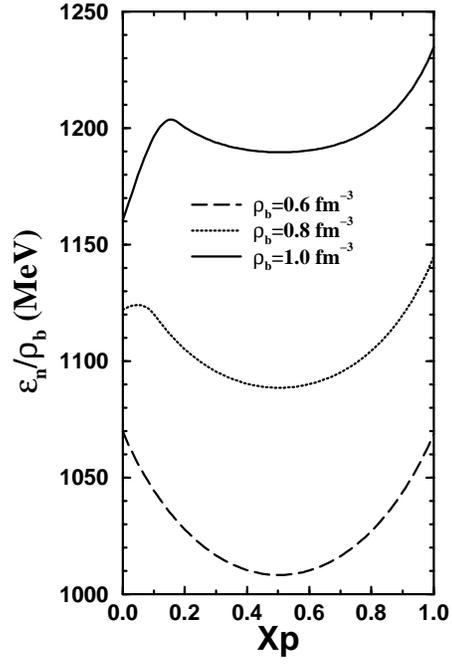}
\end{center}
\caption{
Total energy density $\epsilon_n$ of
 nuclear matter obtained
considering energy minimisation with respect to
spin polarisation parameters are shown as a function
of proton fraction $x_p$ for three particular
values of baryon density $\rho_b$.
}
\label{fig12}
\end{figure}
\newpage
\begin{figure}
\begin{center}
\leavevmode
\epsfxsize = 8cm
\epsffile[40 85 430 710]{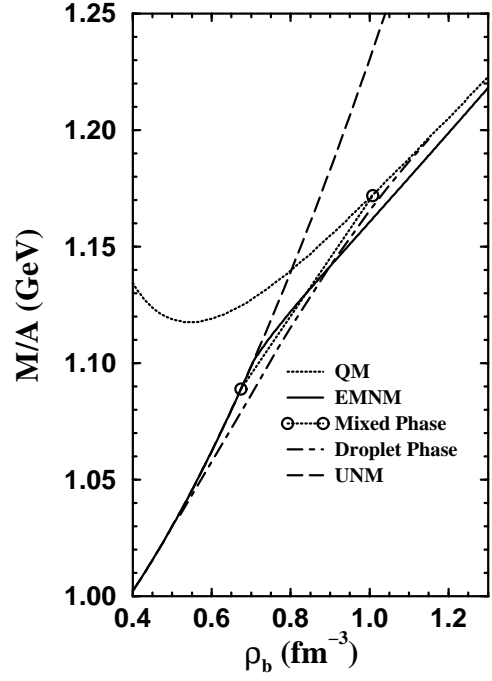}
\end{center}
\caption{
Same as in Fig. 9, but the droplet phase consisting
of unpolarised nuclear matter and strange quark
matter droplets are also shown.
}
\label{fig13}
\end{figure}
\newpage
\begin{figure}
\begin{center}
\leavevmode
\epsfxsize = 8cm
\epsffile[40 85 430 710]{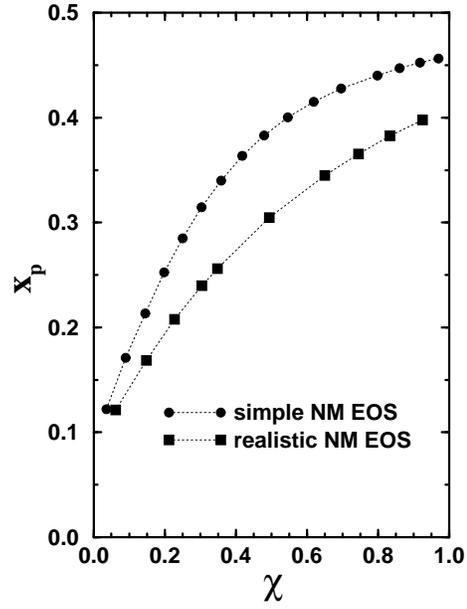}
\end{center}
\caption{
Values of proton fraction $x_p$ obtained for droplet
phase using bag model equation of state (EOS) for quark
matter, and with a simple parametrised EOS 
(\ref{nmeos1}) and a realistic one 
(\ref{msbe},\ref{msbp}) for nuclear matter are
shown as a function of volume fraction $\chi$.
}
\label{fig14}
\end{figure}
\newpage
\begin{figure}
\begin{center}
\leavevmode
\epsfxsize = 8cm
\epsffile[40 85 430 710]{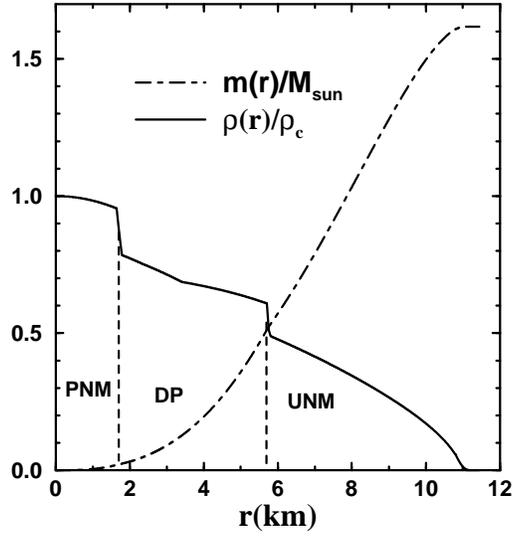}
\end{center}
\caption{
Baryon density $[\rho (r)]$ and mass $[m(r)]$ distributions
corresponding to the maximum mass configuration
obtained 
with bag model (\ref{bageos}
 and realistic nuclear matter (\ref{msbe},\ref{msbp})
equations of state are shown
as a function of the radial distance $r$.
$M_{\rm sun}(\equiv M_{\odot })$ is the solar mass and
$\rho_c$ is the baryon density at the core $(r=0)$ of
the star. (For details see section VI.)
}
\label{fig15}
\end{figure}
\newpage
\begin{figure}
\begin{center}
\leavevmode
\epsfxsize = 8cm
\epsffile[40 85 430 710]{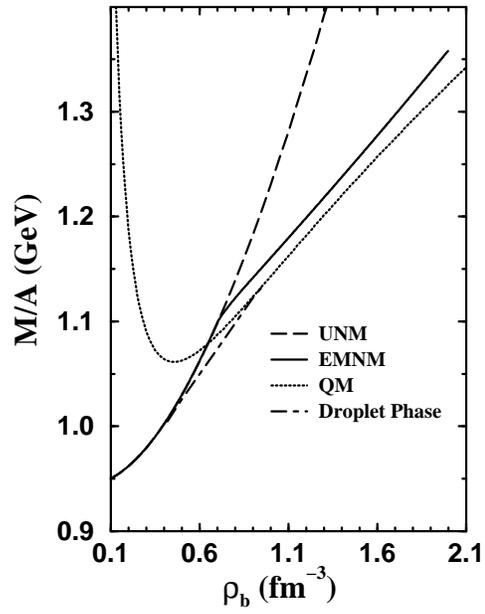}
\end{center}
\caption{
Same as in Fig. 9, but with $B^{1/4} = 170 \ MeV$.
}
\label{fig16}
\end{figure}
\newpage
\begin{figure}
\begin{center}
\leavevmode
\epsfxsize = 8cm
\epsffile[40 85 430 710]{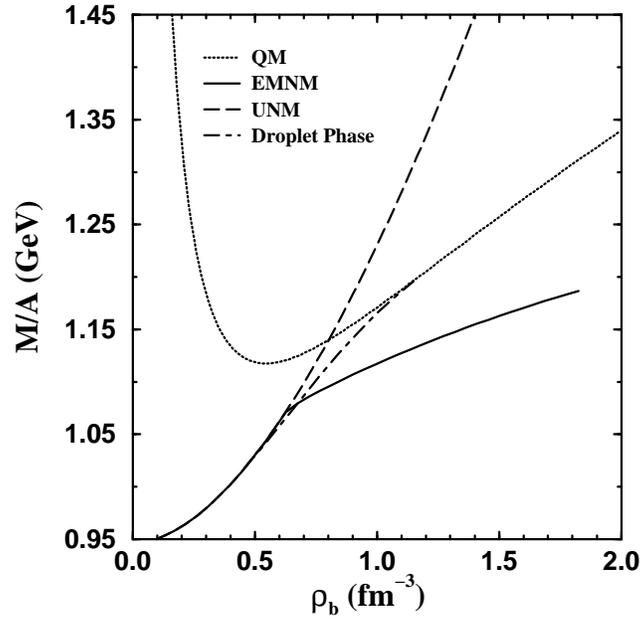}
\end{center}
\caption{
Same as in Fig. 9, but with $E_Y=17\ MeV$.
}
\label{fig17}
\end{figure}
\newpage
\begin{table}
\begin{tabular}{cccccc}
$n$ & $C$ & {${(M/A)}_{P=0}$} & {$\rho_c/\rho_o$} &
{$R$} & {$M_{max}/M_{\odot}$} \\
{} & {$(MeV\ fm^{-3n})$} & {$(MeV)$} & {} & {$(km)$} & {} \\
\hline
  & 127 & 860.0 & 7.31 & 10.5 & 1.85 \\
${1\over 3}$ & 137 & 893.3 & 8.20 & 9.7 & 1.71 \\
 	     & 147 & 925.3 & 9.10 & 9.1 & 1.60 \\
\hline
 &85 & 851.6 & 8.25 & 9.8 & 1.86 \\
${2\over 3}$             & 95 & 883.8 & 9.22 & 9.1 & 1.73 \\
	    & 105 & 913.8 & 10.18 & 8.5 & 1.62 \\
\hline
 & 65 & 855.6 & 8.72 & 9.4 & 1.86 \\
1 & 75 & 886.8 & 9.71 & 8.8 & 1.73 \\
  & 85 & 914.9 & 10.66 & 8.2 & 1.62 \\
\end{tabular}
\caption {
Values of mass per baryon
of flavour symmetric strange quark matter
at zero pressure${(M/A)}_{P=0}$, and
central density $\rho_c$, radius $R$ and mass $M_{\rm max}$
corresponding to the maximum mass configuration of the
strange quark star are given for few sets of the mass
parameters $(n,C)$.
$\rho_o$ is the normal nuclear
matter density.
}
\end{table}

\begin{thebibliography}{99}

\bibitem{hewish}
A. Hewish, S.J. Bell, J.D.H. Pilkington,
P.F. Scott, and R.A. Collins, Nature {\bf 217} (1968) 709.

\bibitem{gold}
T. Gold, Nature {\bf 218} (1968) 731; {\bf 221} (1969) 25.

\bibitem{prakash} 
M. Prakash, {\it Nuclear equation of state }, Eds.
A. Ansari and L. Satpathy, (World Scientific, Singapore, 1996).

\bibitem{shapiro} 
S.L. Shapiro and S.A. Teukolsky, Black holes, white dwarfs
and neutron stars (Wiley, New York, 1983).

\bibitem{itoh} 
N. Itoh, Prog. Theor. Phys. {\bf 44} (1970) 291.

\bibitem{brecher} 
K. Brecher and G. Caporaso, Nature {\bf 259} (1976) 378.

\bibitem{baym} 
G. Baym and S. Chin, Phys.Lett.{\bf 62B} (1976) 241.

\bibitem{larry} 
B. Freedman and L. McLerran, Phys.Rev.{\bf D17} (1978) 1109.

\bibitem{alcock} 
C. Alcock, E. Farhi, and A. Olinto, Astrophys.J. {\bf 310}
(1986) 261.

\bibitem{haensel} 
P. Haensel, J.L. Zdunik, and R. Schaeffer,
Astron. Astrophys. {\bf 160} (1986) 121.

\bibitem{nkg} 
N.K. Glendenning, Phys.Rev.Lett.{\bf 63} (1987) 1825.

\bibitem{witten} 
E. Witten, Phys.Rev.{\bf D30} (1984) 272.

\bibitem{li} 
X. D. Li, Z. G. Dai, and Z. R. Wang, Astron. Astrophys.
{\bf 303 } (1995) L1-L4.

\bibitem{madsen} 
J. Madsen, Astro-ph 9601129.

\bibitem{glen} 
N.K. Glendenning, Phys.Rev.{\bf D46} (1992) 1274;
Phys.Rep. {\bf 264} (1996) 143.

\bibitem{heisel} 
H. Heiselberg, C.J. Pethick, and E.F. Staubo, Phys. Rev.
lett. {\bf 70} (1993) 1355.

\bibitem{drago} 
A. Drago, U. Tambini, and M. Hjorth-Jensen, Phys.Lett.
{\bf B380} (1996) 13.

\bibitem{Um97} 
V.S. Uma Maheswari, D.N. Basu, J.N. De, and S.K. Samaddar, 
Nucl. Phys. {\bf A 615} (1997) 516.

\bibitem{chodos} 
A. Chodos, R.L. Jaffe, K. Johnson, C.B. Thorn and V.F.
Weisskopf, Phys.Rev.{\bf D9} (1974) 3471.

\bibitem{olive} 
K.A. Olive, Nucl.Phys.{\bf B190} (1981) 483.

\bibitem{fowler} 
G.N. Fowler, S. Raha, and R.M. Weiner, Z.Phys.{\bf C9}
(1981) 271.

\bibitem{boal} 
D.H. Boal, J. Schachter and R.M. Woloshyn,
Phys.Rev. {\bf D26} (1982) 3245.

\bibitem{chakra} 
S. Chakrabarty, Phys.Rev.{\bf D43} (1991) 627;
Nuovo Cimento {\bf 106B} (1991) 1023.

\bibitem{ben} 
O.G. Benvenuto and G. Lugones, Phys.Rev.{\bf D51}
(1995) 1989.

\bibitem{glenf} 
M.B. Christiansen and N.K. Glendenning, ASTRO-PH/9706056.

\bibitem{Um96} 
V.S. Uma Maheswari, Phys. Rev. {\bf D\ 54 } (1996) 3389.

\bibitem{hooft} 
G. tHooft, Nucl.Phys.{\bf B72} (1974) 461;
Phys.Rev.{\bf D14}
(1976) 3432.

\bibitem{tegen} 
R. Tegen, Ann.Phys.{\bf 197} (1990) 439.

\bibitem{nambu} 
Y. Nambu and G. Iona-Lasinio, Phys.Rev.{\bf 122}
(1961) 345; {\bf 124} (1961) 246.

\bibitem{parija} 
B.C. Parija, Phys.Rev.{\bf C45} (1992) 415.

\bibitem{bohr} 
A. Bohr and B. Mottelson, {\it Nuclear Structure },
(Benjamin Inc., Amsterdam), (1969).

\bibitem{db} 
D. Bandyopadhyay, C. Samanta, S.K. Samaddar and J.N. De,
Nucl. Phys.{\bf A511}(1990) 1.

\bibitem{hugen} 
N.M. Hugenholtz and L. Van Hove, Physica(Utrecht)
{\bf 24} (1958) 363.

\bibitem{ph}
P. Haensel, Z.Phys.{\bf A274} (1975) 377.

\bibitem{arh} 
M. Abd-Alla, S. Ramadan and M.Y. Hassan, Phys.Rev.{\bf
C36}(1987) 1565.

\bibitem{ub} 
H. Uberall, Electron scattering from complex nuclei, (N.Y.)
, Academic Press (1971).

\bibitem{mmm} 
M.M. Majumdar, S.K. Samaddar, N. Rudra and J.N. De,
 Phys.Rev.{\bf C49} (1994) 541.

\bibitem{fp} 
B. Friedman and V.R. Pandharipande, Nucl.Phys.{\bf A361}
(1981) 502.

\bibitem{rbw} 
R.B. Wiringa, V. Fiks and A. Fabrocini, Phys.Rev.{\bf C38}
(1988) 1010.

\bibitem{fmt}R.P. Feynman, N. Metropolis and E. Teller, Phys.Rev.{\bf
75}(1949) 1561.

\bibitem{bps}G. Baym, C. Pethick and P. Sutherland, Astrophys.J.{\bf
170}(1971) 299.

\bibitem{bbp} G. Baym, H.A. Bethe and C.J. Pethick, Nucl.Phys.{\bf A175}
(1971) 225.

\end{thebibliography}
\end{document}